\documentclass[pdflatex,sn-vancouver-num,iicol]{sn-jnl}

\usepackage{graphicx}%
\usepackage{multirow}%
\usepackage{amsmath,amssymb,amsfonts}%
\usepackage{amsthm}%
\usepackage{mathrsfs}%
\usepackage[title]{appendix}%
\usepackage{xcolor}%
\usepackage{textcomp}%
\usepackage{manyfoot}%
\usepackage{booktabs}%
\usepackage{algorithm}%
\usepackage{algorithmicx}%
\usepackage{algpseudocode}%
\usepackage{listings}%

\usepackage{diagbox}
\usepackage{float}
\usepackage[acronym]{glossaries}%
\usepackage{comment}%
\usepackage{array}     
\usepackage{subcaption}

\hypersetup{
  colorlinks=true,
  linkcolor=black,
  citecolor=black,
  urlcolor=black
}

\newacronym{fhe}{FHE}{Fully Homomorphic Encryption}
\newacronym{phe}{PHE}{Partial Homomorphic Encryption}
\newacronym{he}{HE}{Homomorphic Encryption}
\newacronym{swhe}{SWHE}{Somewhat Homomorphic Encryption}
\newacronym{simd}{SIMD}{Single Instruction/Multiple Data}
\newacronym{dp}{DP}{Differential Privacy}
\newacronym{smc}{SMC}{Secure Multi-party Computation}
\newacronym{ppml}{PPML}{privacy-preserving machine learning}
\newacronym{nb}{NB}{Naive Bayes}
\newacronym{sift}{SIFT}{scale-invariant feature transform}
\newacronym{svm}{SVM}{Support Vector Machine}
\newacronym{ai}{AI}{Artificial Intelligence}
\newacronym{iot}{IoT}{Internet of Things}
\newacronym{iomt}{IoMT}{Internet of Medical Things}
\newacronym{ldp}{LDP}{Local Differential Privacy}
\newacronym{gdp}{gdp}{Global Differential Privacy}
\newacronym{qos}{QoS}{Quality of Service}
\newacronym{ei}{EI}{Edge Intelligence}
\newacronym[plural=EHRs,firstplural=Electronic Health Records (EHRs)]{ehr}{EHR}{Electronic Health Record}
\newacronym{fl}{FL}{Federated Learning}
\newacronym[plural=SSSs,firstplural=Secret-Sharing Schemes (SSSs)]{sss}{SSS}{Secret-Sharing Scheme}
\newacronym[plural=DNNs,firstplural=Deep Neural Networks (DNNs)]{dnn}{DNN}{Deep Neural Network}
\newacronym{rdp}{RDP}{Rényi Differential Privacy}
\newacronym{ma}{MA}{Moments Accountant}

\makeglossaries

\begin{document}

\title[A Privacy-Preserving Machine Learning Framework for Edge Intelligence: An Empirical Analysis]{A Privacy-Preserving Machine Learning Framework for Edge Intelligence: An Empirical Analysis}








\author*[1]{\fnm{Quoc Lap} \sur{Trieu}}\email{19263045@student.westernsydney.edu.au}

\author[1]{\fnm{Bahman} \sur{Javadi}}
\email{b.javadi@westernsydney.edu.au}

\author[1]{\fnm{Jim} \sur{Basilakis}}
\email{j.basilakis@westernsydney.edu.au}

\affil[1]{%
  \orgname{Western Sydney University},
  \orgdiv{School of Computer, Data and Mathematical Sciences},
  \orgaddress{%
    \city{Sydney},
    \state{NSW},
    \country{Australia}%
  }%
}


\abstract{As \acrfull{ei} becomes increasingly prevalent in domains such as smart healthcare, manufacturing, and critical infrastructure, ensuring data privacy while maintaining system efficiency is a growing challenge. This paper presents a new \acrfull{ppml} framework tailored
for \gls{ei} applications, including a four-layer system architecture and training and inference algorithms. We focus on three leading approaches: \acrfull{dp}, \acrfull{smc}, and \acrfull{fhe}, and assess their impact on key performance metrics, including model accuracy, response time, and energy consumption. Results from real implementation and extensive trace-based simulations of inference tasks show that \gls{dp} generally preserves throughput and latency close to plaintext baselines, while accuracy drops with model complexity (up to 35\% on AlexNet and under 18\% on LeNet for FordA). \gls{smc} performance is driven by communication; network bandwidth and round complexity determine end-to-end latency. For AlexNet, increasing link capacity from 250 Mbps to 500 Mbps reduces latency by about 30\%. \gls{fhe} is highly sensitive to model structure and numerical precision (bit width), with tighter parameters imposing substantial compute overhead; we observe roughly a 1000\(\times\) increase in response time compared to \gls{dp}. Beyond efficiency, \gls{dp} shifts the privacy-utility-extractability frontier by reducing the attacker’s data efficiency in black-box model stealing, whereas \gls{smc} and \gls{fhe}, while protecting inputs and parameters during inference, require complementary output controls to achieve similar resistance to extraction.
These findings provide critical insights into the trade-offs between privacy, performance, and resource efficiency in edge computing scenarios.

}

\keywords{Fully Homomorphic Encryption, Differential Privacy, Secure Multiparty Computation, Edge Intelligence Applications, Privacy-Preserving Machine Learning}



\maketitle

\section{Introduction}\label{sec1}



Edge intelligence (EI) is driving advancements in applications such as disease diagnosis in smart healthcare, data acquisition in smart grids, and the management of critical infrastructure \cite{RN262}. By combining edge computing with artificial intelligence, \gls{ei} enables local devices, such as smartphones, to perform real-time data processing and autonomous decision-making, even under resource constraints \cite{RN221, RN261, altintas2024streamlined}.  However, the widespread deployment of \gls{ei} brings significant challenges in safeguarding user privacy and securing sensitive data \cite{singh2023edge}.

Recognizing the critical importance of data privacy in modern applications, many developers have decided to leverage privacy-preserving techniques to protect the privacy of users' sensitive data while performing machine learning tasks~\cite{mercier2021evaluating}. In edge-centric deployments, however, privacy and security requirements extend beyond protecting data at rest: they must hold end-to-end across sensing, on-device/edge processing, transmission, and cloud-side analytics. This is especially acute in remote health monitoring, where highly sensitive medical records and continuous patient streams must remain confidential and tamper-resistant as they are accessed and exchanged among stakeholders, motivating decentralized architectures that combine blockchain-backed auditability with identity and access-control mechanisms to reduce reliance on a single trusted party~\cite{seidi2025securing}. Meanwhile, edge-IoT environments are routinely exposed to confidentiality- and availability-threatening attacks (e.g., man-in-the-middle, injection, and distributed denial of service), which have driven the adoption of data-driven defenses—such as deep-learning-based intrusion detection frameworks deployed at the edge for timely detection and response~\cite{boubertakh2025hymd2i}.

However, choosing an appropriate privacy scheme remains non-trivial, particularly at the edge where compute and memory resources are constrained and latency budgets are tight~\cite{ding2022privacy}. In other words, with the limitation of edge computing on processing power and memory capacity, a comparative evaluation of different privacy-preserving schemes is essential for developing effective \acrfull{ppml} techniques at the edge. 

In this paper, we perform an extensive evaluation of different privacy-preserving techniques in the context of smart applications at the edge. The methodology provides a comparative analysis of three widely used techniques, namely, \acrfull{dp}, \acrfull{smc}, and \acrfull{fhe} \cite{mercier2021evaluating, RN220}. Basically, \gls{dp} ensures to preserve users' privacy by adding a certain noise to the sensitive data during training while keeping the trade-off between the level of model accuracy and data privacy \cite{RN250}. The second technique, \gls{smc}, allows multiple parties to perform joint computations without sharing data. In other words, the protocol preserves each party's data privacy but can still perform the operation to get the result \cite{RN241}. The final technique \gls{fhe} is one of the main types of homomorphic encryption, a cryptography-based technique that allows computations on encrypted data without the presence of private keys. 

To evaluate the performance of these three privacy-preserving techniques, we conduct experiments using both \textit{real implementation} and \textit{trace-based simulations} on different deep-learning models and time-series datasets. The primary task in the experiments is machine learning \textit{inference}, which is the most common operation in \gls{ei} systems. The results are then analyzed across several metrics, including accuracy, response time and energy consumption. To the best of our knowledge, this is the first study to conduct a comprehensive performance evaluation of privacy-preserving techniques for \gls{ei} applications. The main contributions of this work are:
\begin{itemize}
    \item The development of a new \gls{ppml} framework tailored for \gls{ei} applications, structured into four hierarchical layers---Cloud, Edge Server, Edge Device, and Sensor---and incorporating four core components: Data, Model, System, and Application.
    \item Implementation and evaluation of  \gls{ppml} techniques on heterogeneous edge devices to develop a realistic dataset for performance evaluation.
    \item A trace-based simulation for evaluating \gls{ppml} techniques using real-world data collected from diverse datasets and edge devices under different workloads.
    \item A comprehensive security analysis and performance evaluation of \gls{ppml} techniques for various \gls{ei} application domains, assessing accuracy, response time and energy consumption.
\end{itemize}





The rest of the paper is organized as follows. Section~\ref{section:background} reviews related work on \gls{ppml} performance evaluation. Section~\ref{section:framework} describes the architecture of the proposed framework and its evaluation methodology.  Section~\ref{section:evaluation} then details the development of the \gls{ppml} framework, including the training and inference algorithms, the experimental setup, and the trace collection process. Section~\ref{section:results} reports and analyzes the evaluation results.
Section~\ref{section:security-analysis} presents a comparative security analysis of the considered privacy-preserving techniques. Finally, Section~\ref{section:conclusion} summarizes the key findings and concludes the paper.
    
\section{Related Work}
\label{section:background}
In the following section, we discuss existing research on privacy preservation approaches in edge intelligence and broader machine learning frameworks.

\subsection{Privacy-Preserving Techniques for Edge Intelligence}

Recently, privacy-preserving techniques at the edge have gained significant attention for protecting sensitive user data and ensuring compliance \cite{RN268, RN269}. This section reviews recent studies on implementing \gls{ppml} applications using various privacy-preserving schemes in \gls{ei} systems.

\subsubsection{Differential Privacy}

Introduced in 2006, \acrfull{dp} is a key privacy technique \cite{RN250}. It enables control over the noise added to data, using Laplace, Gaussian, or Exponential mechanisms based on risk and data traits. \gls{dp} is classified by noise location: \gls{ldp}, where noise is added locally before storage, and \gls{gdp}, where it is applied to the database post-storage \cite{RN250, RN251}.

Many studies in edge computing apply \acrfull{dp} for \gls{fl} to preserve privacy during AI model training. Wan et al. \cite{RN202} proposed integrating blockchain-enabled \gls{fl} with \gls{dp} to balance privacy and data utility. Another approach \cite{RN214} modifies the training method to optimize learning rates while adding noise to the model updates. Studies by Xu et al. \cite{RN204} and Kai et al. \cite{RN210} focus on enhancing privacy in blockchain-enabled frameworks. Xu et al. \cite{RN204} specifically address protecting patient data in coronary heart disease diagnosis using \gls{ldp} without a trusted entity. 

In \cite{RN210}, the authors used the Laplace mechanism to protect the privacy of edge locations in the strategy for task allocations. As one of the big tech companies famous for privacy protection, Apple has been working in \gls{dp} to enhance the level of data privacy while being able to collect more data. In \cite{RN205}, the team proposed a system leveraging \gls{ldp} to enable the ability to train \gls{ai} models at large-scale data while preserving their privacy.

Regarding the application of \gls{dp} in computer vision, Mao, Yunlong, et al. \cite{mao2018privacy} proposed a secure solution for training the \gls{dnn} face recognition model. They also applied an offloading strategy to split the \gls{dnn} among user devices and the edge server while preserving the privacy of both data and model parameters. Another approach to protecting the data privacy of the machine learning models in edge computing systems is from \cite{RN253}. The approach first partitions the data into blocks and distributes it to edge nodes for training and processing; it then aggregates the entire dataset for further processing. During the whole process, the authors used the Laplace mechanism to add noise to both data blocks and the entire dataset.

\subsubsection{Secure Multi-party Computation}

According to Feng and Yang \cite{RN241}, \gls{smc} protocols are typically built from two core primitives: secret sharing and garbled circuits. Secret sharing protects private inputs by distributing shares of the secret, while garbled circuits enable computation over encrypted values. \glspl{sss} are commonly grouped into three categories: Linear, Additive, and Shamir, distinguished by how they generate, distribute, and reconstruct shares.

Several works integrate \gls{smc} into \gls{fl} to provide privacy-preserving training. Li et al. \cite{RN216} and Kanagavelu et al. \cite{RN201} both adopt \gls{smc}-based mechanisms, with Kanagavelu et al.\ focusing on overhead and scalability, and Li et al.\ employing Shamir’s scheme for vehicular fog computing. Other designs rely solely on secret sharing, such as Liu et al. \cite{RN217} for secure data verification in smart grids, or Praveen et al. \cite{praveen2024design} for searchable encryption. Olakanmi and Odeyemi \cite{RN209} further extend \gls{sss} by distributing computation across multiple workers and embedding incentive and trust models to ensure fairness.

\gls{he} is another key building block frequently combined with \gls{smc} in practical systems. In vehicular networking, \cite{RN200} proposes a distributed computation offloading scheme that reduces the processing load on vehicles and roadside units while enhancing model security through the joint use of \gls{he} and \gls{smc}. Khan et al. \cite{RN207} design a secure data aggregation protocol for smart grids that leverages Shamir’s secret sharing to transmit data to different nodes and Paillier \gls{he} for aggregation. For \gls{iomt}, \cite{RN188} introduces a composite framework that encrypts medical plaintext with \gls{he} and employs \gls{sss} to distribute computation across nodes. To mitigate relay and delay attacks during data transmission, the framework also incorporates a secure time-synchronization protocol from \cite{RN271}.

\subsubsection{Homomorphic Encryption}

Homomorphic encryption is used for protecting data privacy in scenarios like outsourcing sensitive data to the cloud \cite{cheon2022introduction}. There are three types: partial homomorphic encryption (PHE), levelled homomorphic encryption (LHE), and \acrfull{fhe}. 
PHE supports repeated addition or multiplication, with Paillier and RSA as common examples. LHE extends this to both operations but with limited depth; BFV and CKKS are popular LHE schemes used in encrypted machine learning inference \cite{RN220}. \gls{fhe} supports an unlimited number of operations and is the most powerful. Among \gls{fhe} schemes, TFHE excels at fast Boolean gate-level operations, making it ideal for encrypted logic, while the Brakerski-Gentry-Vaikuntanathan (BGV) scheme supports both leveled and fully homomorphic operations with efficient noise management \cite{RN220}. In smart healthcare applications, Alabdulatif et al. \cite{RN192} used \gls{fhe} to enhance privacy in unsupervised clustering techniques like K-Means and Fuzzy C-Means. They also identified the BGV scheme as promising for privacy-preserving applications. Another medical \gls{ppml} approach \cite{RN189} combines homomorphic encryption, blockchain, and \gls{fl} for privacy protection.

Around 2020, amid the COVID-19 breakout, it was important to prevent and detect early cases based on data analysis from \gls{iomt} applications while ensuring the privacy of sensitive patient-related data. The authors of \cite{RN190} presented a strategy named privacy-enhanced data fusion, applying Paillier \gls{he} as one of the main four components in the approach. The authors of \cite{RN273} also employed \gls{fhe} and the blockchain technology underpinning a privacy-preserving deep learning model for Vehicular Ad-hoc Networks (VANETs) to protect data privacy.

In an attempt to apply a privacy-preserving solution at the edge, the author of \cite{catalfamo2022homomorphic} tested the performance of different \gls{he} techniques by using two libraries: Python-Paillier and PyFhel. However, the approach only uses an edge layer to encrypt the data; the computation process still relies on the cloud layer. J Hrzich et al. also showed an experiment on applying different \gls{he} techniques for several ML models in \cite{hrzich2022experimental}. However, the research considers classification prediction models with only a few \gls{he} techniques. The research on \cite{xu2022edge} applied \gls{phe} as a solution for a secure blockchain application. The application relies \gls{phe} to execute the verification process based on the encrypted value at the blockchain nodes.

\subsection{Privacy-Preserving Machine Learning Evaluation Framework}

There is limited work on privacy-preserving machine learning evaluation frameworks in edge environments. A privacy-preserving task allocation framework for edge computing based mobile crowdsensing is introduced by Ding et al. \cite{ding2022privacy}, which ensures user data confidentiality while efficiently assigning tasks. Using \gls{he} and \gls{dp}, the framework achieves a balance between privacy, computational efficiency, and allocation accuracy. However, while the study compares the efficiency of these techniques across varying numbers of tasks and users, it does not address the performance evaluation of machine learning tasks within the framework. 
Shokri et al. \cite{shokri2017membership} introduced a framework using membership inference attacks to assess ML privacy risks. Shafee et al. \cite{shafee2023false} extended this to detect anomalies in electric vehicle charging systems with both membership inference and model inversion attacks. While these frameworks lay the groundwork for assessing the robustness of privacy-preserving techniques, these works do not compare different \gls{ppml} methods or their practical use in distributed or edge environments.
Mercier et al. \cite{mercier2021evaluating} conducted an evaluation of various privacy-preserving techniques for machine learning, with a focus on time-series classification. The paper specifically evaluates the applicability of \gls{dp}, \gls{fl}, and \gls{smc} on time-series datasets. While it provides a comprehensive analysis of accuracy across different techniques and datasets, it overlooks inference performance and the impact on energy consumption. Additionally, the evaluation is centred on cloud computing environments, with limited consideration of edge computing scenarios.

These studies underscore the increasing focus on privacy-preserving techniques for edge computing and IoT systems. Shafee et al. \cite{shafee2025privacy} highlight ongoing challenges in evaluating these techniques, particularly due to real-world complexities. Notably, gaps remain in assessing machine learning performance, inference efficiency, and energy consumption within edge environments. Many \gls{ppml} methods applied at the edge use specific privacy techniques or their combinations, often without a thorough understanding of their performance trade-offs. This lack of systematic evaluation impedes informed choices when selecting methods for diverse applications. Therefore, comprehensive comparative evaluations are essential to guide users in choosing the most suitable privacy-preserving techniques for their specific needs. In this study, we did not consider any \gls{fl} techniques as they primarily focus on training, whereas our emphasis is on \textit{inference}, the most common operation in \gls{ei} applications.

\section{System Architecture}
\label{section:framework}
In this section, the system architecture and implementation tools for privacy-preserving edge intelligence will be explained. 
\subsection{Architecture}


We first define four primary components of the \gls{ei} system: Data, Model, System, and Application.

\paragraph{Data component}
In general, the Data component involves all the operations related to data and the data itself. In other words, the Data component could include all data generated from real-time data collection of sensor devices, from database import operations or from user's input of the top-level applications. The Data component also covers all data manipulations, such as data quantization, encoding, or data encryption. 

\paragraph{Model component}
As the core component of an \gls{ei} system that extends the ability of edge computing, the Model component decides how \gls{ai} models are trained and deployed. It is related to machine learning models and systematic ways of learning and inference processes.

\paragraph{System component}
The System component represents the computing infrastructure and properties of the \gls{ei} system, including memory, CPU, edge devices, and their performance including reliability, energy usage, throughput, and latency. 


\paragraph{Application component}
The Application component provides the service to users based on the Data, Model, and System components. It is a component that decides how an application can integrate with the \gls{ei} system. 


\vspace{0.7em}

We believe these components play an essential role in implementing every \gls{ei} application. The data component manages all data-related aspects of \gls{ei} to ensure the quality and consistency of data for processing. Based on the Data component, the Model component ensures the performance of the \gls{ai} models in training and inference. At the same time, using the System component, users can easily define different system properties to ensure they meet the expected performance from the Application component, such as expected latency and throughput.

\begin{figure*}[h]
  \centering \includegraphics[width=0.85\textwidth]{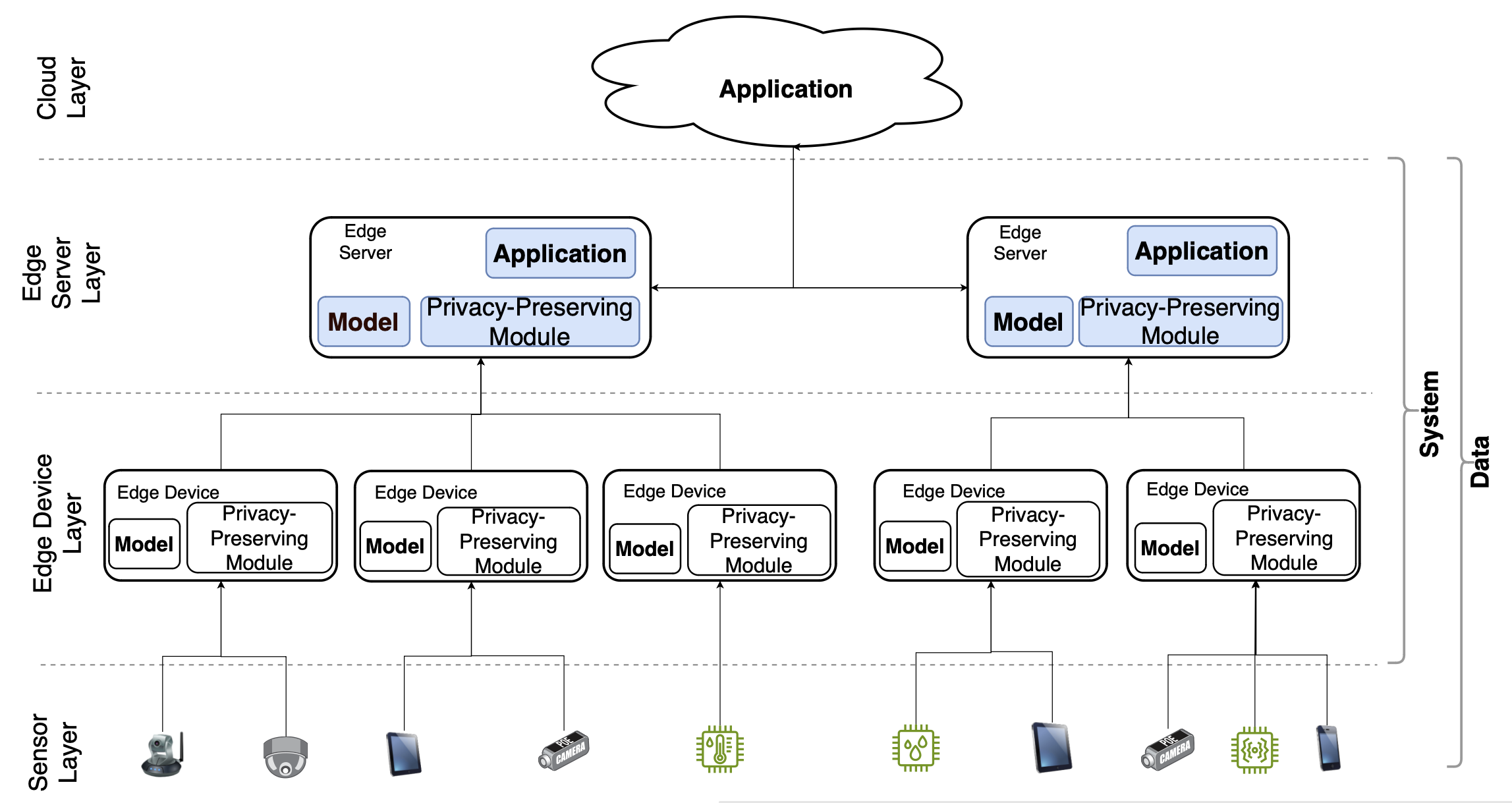}
  \caption{ The Proposed System Architecture with Data, Model, System, and Application Components}
  \label{fig:proposed-architecture}
\end{figure*}

Figure \ref{fig:proposed-architecture} provides an overview of our proposed architecture for privacy-preserving \gls{ei}. The architecture is structured into four layers: the sensor layer, edge device layer, edge server layer, and cloud layer. These layers collectively enable efficient data processing while maintaining privacy and optimizing system performance.
As can be seen in the figure, the four components: Data, Model, System, and Application, closely cooperate to form an \gls{ei} architecture. Our architecture relies on a privacy-preserving module to be developed in the edge server and edge device layer to ensure the application's privacy while executing the machine learning tasks.

The sensor layer consists of IoT devices that collect data such as images, videos, and temperature readings. Once data is gathered, the edge device layer handles processing tasks within its limited computational capacity. By leveraging both the model and the privacy-preserving module, it performs tasks such as data encryption and pre-processing. However, due to the limitations of embedded devices, more resource-intensive tasks are offloaded to the edge server layer, which offers higher processing power. This edge server layer also integrates the model and privacy-preserving module, maintaining privacy while handling more demanding tasks, such as secure inference. Notably, the edge device and edge server layers may use different \gls{ai} models tailored to their respective processing capabilities.
The architecture also includes a cloud server layer to support application use cases, as shown in Figure \ref{fig:proposed-architecture}. Both edge server and cloud layers incorporate the application component, enabling developers to implement and deploy \gls{ppml} applications via either layer, depending on their requirements.

This architecture supports a wide range of privacy-sensitive \gls{ei} applications. In smart healthcare, for instance, patient data collected from wearable devices (e.g., heart rate, ECG) can be encrypted and processed directly in its encrypted form before being transferred to hospital-edge servers for secure inference. This enables real-time diagnosis and patient monitoring while preserving data privacy. In autonomous vehicles, each vehicle functions as an edge device, collecting sensor data and performing secure computations locally. The encrypted results can then be transmitted to edge servers for further processing, ensuring low-latency object detection and navigation without compromising data protection. By adopting a multi-layered processing approach, the proposed architecture effectively balances performance, privacy, and scalability, enabling secure and efficient execution of various \gls{ei} applications.

\subsection{Privacy-Preserving Tools}


We consider various tools for privacy-preserving techniques, edge frameworks and machine learning models during the evaluation process. Even though many tools are available for each privacy technique or framework, we only select recent state-of-the-art \gls{ppml} techniques \cite{mercier2021evaluating}.


We use TensorFlow Privacy, a Google-led open-source library that applies \gls{dp} via differentially private stochastic gradient descent (DP-SGD) \cite{papernot2019machine}. The library provides a noise multiplier parameter representing how much noise is added to the model during training. Adjusting the parameter value also directly impacts the privacy budget, which represents the allowable amount of information leakage about individual data points.

This study utilizes CrypTen, a library developed by Meta specifically designed for \gls{ppml} applications using \gls{smc} to enable secure model training and inference \cite{knott2021crypten}. CrypTen provides a robust framework for implementing privacy-preserving techniques in machine learning pipelines, allowing secure computation across data sources without compromising model accuracy. The library supports encryption protocols like TLS (Transport Layer Security) to ensure encrypted communication. While \gls{smc} is typically used for collaborative training between organizations, in this study, we employ it for secure inference, ensuring data security across edge nodes within an \gls{ei} system.

Regarding the homomorphic encryption technique, we use the Concrete-ML library to implement \gls{ei}. Concrete-ML is an open-source framework implementing \gls{fhe} for \gls{ppml} applications. The framework is developed by Zama and based on \gls{fhe} over the Torus (TFHE) \cite{RN230, chillotti2020concrete}.

\begin{figure*}[h]
  \centering
  \includegraphics[width=0.85\textwidth]{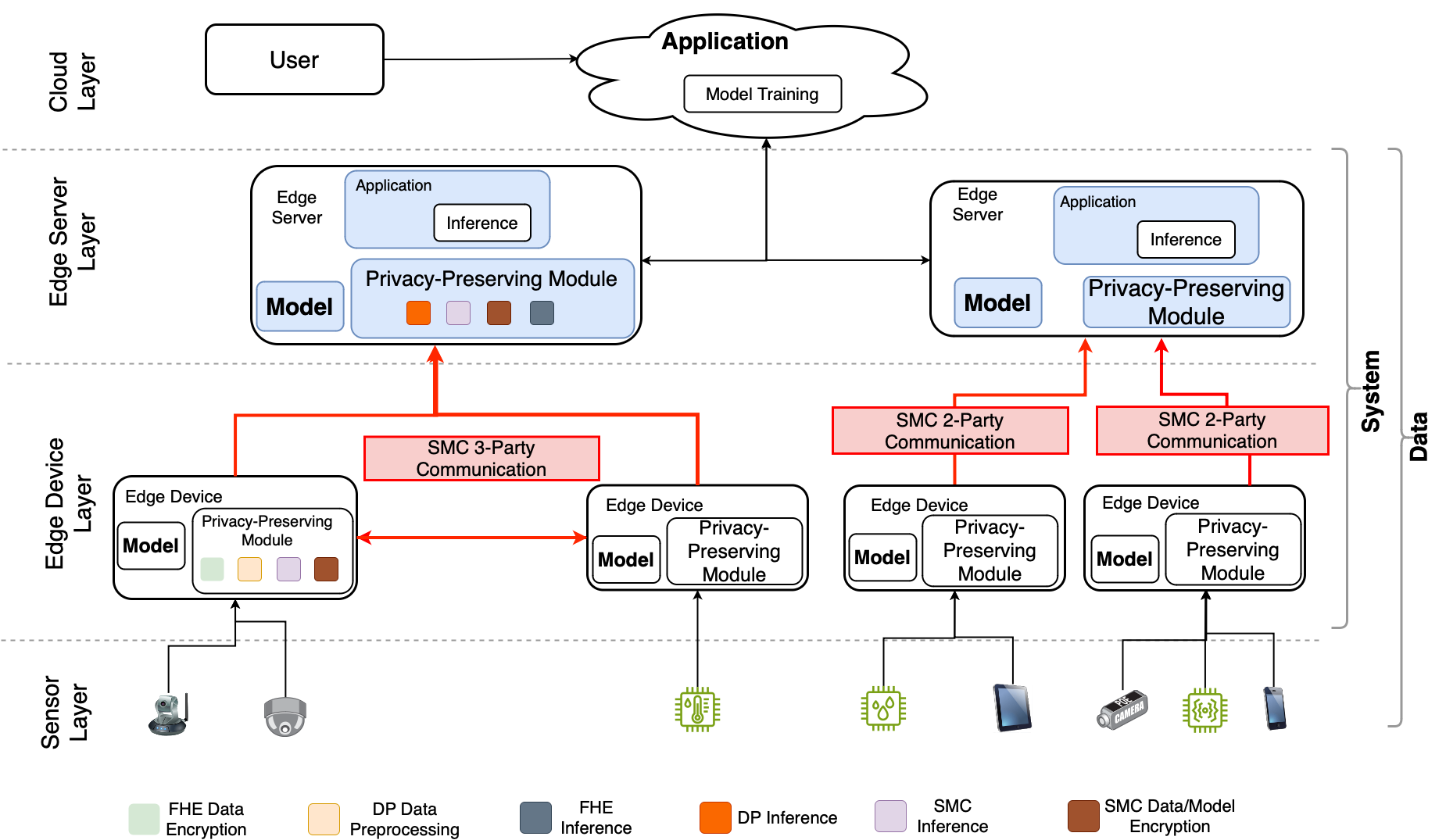}
  \caption{Developed Architecture for Privacy-Preserving Edge Intelligence Framework}
  \label{fig:secure-architecture}
\end{figure*}

\section{The Proposed PPML Framework}
\label{section:evaluation}

This section first introduces the proposed \gls{ppml} framework and then formalizes its workflow by summarizing the cloud-based training/preparation and edge-based inference procedures in two algorithms. It then describes the development process for the proposed framework, starting with the deep learning models and datasets used for evaluating privacy-preserving techniques. Next, it provides an overview of the experimental setup, which is divided into two stages: a hardware experiment and a trace-based simulation. The hardware experiment involves testing AI models with privacy techniques on edge hardware, while the software simulation evaluates system performance and energy efficiency in a large-scale environment using traces from the first experiments. The results are analyzed based on three metrics: inference accuracy, response time, and energy consumption.

Figure~\ref{fig:secure-architecture} depicts the developed architecture used to evaluate three privacy-preserving techniques for \gls{ei} applications across the cloud-edge hierarchy. The cloud layer performs model training, whereas inference is executed within the edge server and edge device layers to reflect latency-sensitive deployment. During operation, the sensor layer continuously generates IoT samples $x(t)$, where $t$ denotes the discrete sample index, and each sample is first received by an edge device for local processing. Depending on the selected privacy method, the edge device either applies data sanitisation or encryption prior to outsourcing inference to the edge server. In particular, under the \gls{fhe} setting the edge device encrypts inputs before transmission so that the edge server can compute predictions directly on ciphertexts, while under the \gls{dp} setting the privacy protection is incorporated during training and inference proceeds on plaintext features using a DP-trained model. In contrast, \gls{smc} distributes the inference computation across the edge server and one or more edge devices in a 2-party or 3-party configuration. In every \gls{smc} configuration, the edge server acts as party~$P_0$ to orchestrate and mediate message exchange, while the remaining parties correspond to edge devices that jointly execute the secure computation. Throughout the experiments, we record both computation time and communication statistics to quantify the runtime cost introduced by each privacy mechanism, with particular emphasis on the communication overhead inherent to \gls{smc}.

\subsection{PPML Training and Inference Algorithms}

Building on the architecture in Figure~\ref{fig:secure-architecture}, we formalize the experimental workflow with two algorithms: a cloud-based model training and deployment preparation, followed by an edge-based privacy-preserving inference and measurement. 

Algorithm~\ref{alg:pp-train-prep} specifies the cloud-based training and preparation stage, producing all artifacts required to execute secured inference at the edge. When $m=\textsf{DP}$, the cloud trains a differentially private model via DP-SGD using TensorFlow-Privacy and deploys the resulting DP-trained model to the edge servers for standard (plaintext) inference. When $m\in\{\textsf{FHE},\textsf{SMC}\}$, the cloud first trains a baseline model. For $m=\textsf{FHE}$, the trained model is compiled into an FHE-executable representation using Concrete-ML in line~8. The associated cryptographic material is then split across layers: the edge device retains the secret key required for decryption, whereas the edge server is provisioned only with the public/evaluation information needed to evaluate the model directly on ciphertexts. During compilation, the precision bit-width $p$ and quantization bit-width $q$ are provided as configuration parameters to control the numerical representation and accuracy-performance trade-off of the compiled model. For $m=\textsf{SMC}$, the baseline model is packaged for CrypTen and deployed to the edge server ($P_0$), enabling secure inference via secret sharing and interactive multi-party computation protocols.

Algorithm~\ref{alg:pp-inference} describes the inference procedure triggered by each incoming sensor sample $x(t)$. Each sample is first received by an edge device, which performs standard local pre-processing before following one of three method-specific branches. Under \gls{dp}, the pre-processed input is forwarded to the edge server, which runs plaintext inference using the DP-trained model and returns the prediction $\hat{y}(t)$. Under \gls{fhe}, the edge device encodes and encrypts the pre-processed input, the edge server evaluates the compiled model directly over encrypted data to produce an encrypted output, and the edge device decrypts the result locally (lines~10--15). Under \gls{smc}, the edge server serves as both model owner and coordinator. It keeps the model weights in plaintext and distributes them to the other parties as secret shares, so that each edge device receives only a share and cannot reconstruct the full model on its own (line~18). The input is secret-shared and inference is executed jointly by the edge server ($P_0$) and the edge devices ($P_{i}$) in 2PC/3PC mode, revealing only the final prediction to the authorised party. Across all settings, we record per-sample computation time and communication cost (e.g., transmitted bytes and round duration) to enable a consistent comparison of privacy overheads, with particular emphasis on the network-dependent overheads of \gls{smc}.

For ease of reference, Table~\ref{tab:algo-notation} summaries the notation used in Algorithms~\ref{alg:pp-train-prep} and~\ref{alg:pp-inference}, including model artifacts, party roles, and timing variables, thereby avoiding repeated symbol definitions in the algorithm text.

\begin{table}[t]
\centering
\scriptsize
\renewcommand{\arraystretch}{1.5} 
\caption{Notation used in Algorithms~\ref{alg:pp-train-prep} and~\ref{alg:pp-inference}.}
\label{tab:algo-notation}
\begin{tabular}{|p{0.20\linewidth}|p{0.70\linewidth}|}
\hline
\textbf{Symbol} & \textbf{Meaning} \\ \hline
$m$ & Selected privacy method: $\{\textsf{DP},\textsf{FHE},\textsf{SMC}\}$ \\ \hline
$x(t)$ & Sensor input at discrete sample index $t$ \\ \hline
$x_{\mathrm{raw}},\, x_{\mathrm{prep}}$ & Raw sample and pre-processed sample at the edge device \\ \hline
$x_{\mathrm{pt}}$ & Encoded plaintext representation of $x_{\mathrm{prep}}$ prior to FHE encryption \\ \hline
$M$ & Baseline (non-private) trained model \\ \hline
$M_{\mathrm{dp}}$ & DP-trained model (trained with DP-SGD) \\ \hline
$\hat{y}(t)$ & Predicted output for sample $x(t)$ \\ \hline
$\varepsilon$ & Privacy budget (smaller $\varepsilon$ implies stronger privacy) \\ \hline
$\delta$ & Failure probability in $(\varepsilon,\delta)$-DP \\ \hline
$C$ & Gradient clipping norm bound in DP-SGD \\ \hline
$\sigma$ & Noise multiplier for Gaussian noise in DP-SGD \\ \hline
$pk,\, sk,\, evk$ & FHE public key, secret key, and evaluation key material \\ \hline
$p$ & Precision bit-width \\ \hline
$q$ & Quantization bit-width \\ \hline
$M_{\mathrm{fhe}}$ & FHE-compiled model representation (Concrete-ML) \\ \hline

$M_{\mathrm{smc}}$ & SMC-encrypted model \\ \hline
$world\_size$ & Number of SMC parties \\ \hline
$P_0$ & Party 0 (edge server; coordinator in \gls{smc}) \\ \hline
$P_{i}$ & Remaining parties (edge devices) in 2PC/3PC \\ \hline
\end{tabular}
\end{table}


\begin{algorithm*}
\caption{Training and Deployment (DP/FHE/SMC)}
\label{alg:pp-train-prep}

\begin{algorithmic}[1]
\Require Privacy method $m \in \{\textsf{DP},\textsf{FHE},\textsf{SMC}\}$; training set $D_{\mathrm{train}}$; DP params $(\varepsilon,\delta,C,\sigma)$; FHE params (calib set $D_{\mathrm{calib}}$, $p$, $q$); SMC setting (2PC/3PC with $P_0$ as EdgeServer).
\Ensure Deployed artifacts for edge inference (model and, if applicable, cryptographic/SMC setup).

\If{$m=\textsf{DP}$}
    \State $M_{\mathrm{dp}} \gets \Call{TrainDP\_SGD}{D_{\mathrm{train}}, \varepsilon,\delta,C,\sigma}$ \# Train DP model using DP-SGD
    \State \Call{DeployToEdgeServer}{$M_{\mathrm{dp}}$} \# Deploy model to edge servers
\Else
    \State $M \gets \Call{TrainModel}{ D_{\mathrm{train}}}$ \# Train baseline model

    \If{$m=\textsf{FHE}$}
        \State \# Compile model for FHE inference (Concrete-ML)
        \State $M_{\mathrm{fhe}} \gets \Call{CompileFHE}{M, D_{\mathrm{calib}}, p, q}$
        \State \# Generate keys and distribute to edge layers
        \State $(sk,pk,evk) \gets {FHE Key Genegation}{}$
    \ElsIf{$m=\textsf{SMC}$}  
        \State \Call{DeployToEdgeServer}{$M$} \# Deploy model to edge servers
        \State \Call{CrypTenInit}{world\_size, rank} \# initialize CrypTen parties (EdgeServer is $P_0$)
    \EndIf
\EndIf
\end{algorithmic}
\end{algorithm*}

\begin{algorithm*}
\caption{Edge Inference and Metric Collection (DP/FHE/SMC)}
\label{alg:pp-inference}

\begin{algorithmic}[1]
\Require Privacy method $m \in \{\textsf{DP},\textsf{FHE},\textsf{SMC}\}$; deployed artifacts from Algorithm~\ref{alg:pp-train-prep}; sensor stream $\{x(t)\}$.
\Ensure Predicted output $\hat{y}(t)$ and logged performance measurements.

\While{\textbf{true}}
    \State $x_{\mathrm{raw}} \gets \Call{SensorRead}{}$
    \State $x_{\mathrm{prep}} \gets \Call{Preprocess}{x_{\mathrm{raw}}}$

    \If{$m=\textsf{DP}$}
        \State \# Transmit the pre-processed sample $x_{\mathrm{prep}}$ to the edge server for inference
        \State \Call{Send}{EdgeDevice$\rightarrow$EdgeServer, $x_{\mathrm{prep}}$}
        \State $\hat{y}(t) \gets \Call{Forward}{M_{\mathrm{dp}}, x_{\mathrm{prep}}}$
        \State \Call{Log}{\textsf{DP}, comm=optional}

    \ElsIf{$m=\textsf{FHE}$}
        \State $x_{\mathrm{pt}} \gets \Call{Encode}{x_{\mathrm{prep}}, \textit{calib\_info}}$
        \State $x_{\mathrm{enc}} \gets \Call{Encrypt}{pk, x_{\mathrm{pt}}}$
        \State \# Transmit the pre-processed sample $x_{\mathrm{prep}}$ to the edge server for inference
        \State \Call{Send}
        {EdgeDevice$\rightarrow$EdgeServer, $x_{\mathrm{enc}}$}
        \State $\hat{y}_{\mathrm{enc}}(t) \gets \Call{FHEForward}{M_{\mathrm{fhe}}, x_{\mathrm{enc}}, evk}$
        \State $\hat{y}(t) \gets \Call{Decrypt}{sk, \hat{y}_{\mathrm{enc}}(t)}$  \hspace{1cm} \# Decrypt $\hat{y}_{\mathrm{enc}}(t)$ with $sk$
        \State \Call{Log}{\textsf{FHE}, comm=bytes/time}

    \Else
        \State $M_{\mathrm{smc}} \gets \Call{Encrypt}{{M}, source=0}$   \# Encrypt model on $P_0$ and secret share with all parties
        \State Select input owner $P_i$ (EdgeDevice)
        \State $x_{\mathrm{smc}} \gets \Call{Encrypt}{x_{\mathrm{prep}}, source=i}$
        \State $\hat{y}_{\mathrm{mpc}}(t) \gets \Call{SMCForward}{M_{\mathrm{smc}}, x_{\mathrm{mpc}}}$
        \State $\hat{y}(t) \gets \Call{Reveal}{\hat{y}_{\mathrm{mpc}}(t), destination=i}$ \hspace{1cm} \# Decrypt $\hat{y}_{\mathrm{mpc}}(t)$
        \State $\textit{stats} \gets \Call{CrypTenCommStats}{}$
        \State \Call{Log}{\textsf{SMC}, bytes=\textit{stats}}
    \EndIf
\EndWhile
\end{algorithmic}
\end{algorithm*}

\subsection{Development of the PPML Framework}

This section describes how the proposed \gls{ppml} framework is realized in practice, covering both the edge hardware implementation and the trace-driven simulation environment used for large-scale evaluation.

\subsubsection{Deep Learning Models and Datasets}
This section presents the main deep-learning models and datasets used for evaluation. The experiments were conducted on three common models with varying layers and inputs, which have been used in similar studies~\cite{mercier2021evaluating,wang2025optimizing}. LeNet-5, a well-known model in the LeNet family, includes two convolutional layers and is commonly used for digit recognition tasks. SqueezeNet is a compact architecture designed to achieve AlexNet-level accuracy with significantly fewer parameters, leveraging Fire modules to reduce model size while preserving performance. AlexNet, introduced by Krizhevsky et al. \cite{Krizhevsky2012}, consists of five convolutional layers and three fully connected layers.

\begin{table*}[t]
\centering
\scriptsize
\renewcommand{\arraystretch}{1.5} 
\caption{Time-series Datasets}
\label{table:uea_ucr_datasets}
\begin{tabular}{|>{\raggedright\arraybackslash}p{3.5cm}
                |>{\raggedright\arraybackslash}p{3.5cm}
                |>{\raggedright\arraybackslash}p{1.5cm}
                |>{\raggedright\arraybackslash}p{1.5cm}
                |>{\raggedright\arraybackslash}p{1.5cm}
                |>{\raggedright\arraybackslash}p{1.5cm}|}
\hline
\textbf{Dataset} & \textbf{Sector} & \textbf{Train} & \textbf{Test} & \textbf{Length} & \textbf{Classes} \\ \hline
FordA & Critical manufacturing & 3601 & 1320 & 500 & 2 \\ \hline
ElectricDevices & Energy & 8926 & 7711 & 96 & 7 \\ \hline
ECG5000 & Public health & 500 & 4500 & 140 & 5 \\ \hline
\end{tabular}
\end{table*}

To evaluate the effectiveness of privacy-preserving techniques in smart applications, three real-world datasets from the UEA \& UCR Time Series Classification Archive are used \cite{Bagnall2021}. Key dataset characteristics, such as the number of training and testing samples and input length, are summarized in Table \ref{table:uea_ucr_datasets}. For the smart healthcare scenario, the ECG5000 dataset is employed, which contains 5,000 heartbeat samples from a patient with severe congestive heart failure. In the smart grid application, the ElectricDevices dataset provides energy consumption data from 251 households, classified into seven categories based on device usage patterns, representing diverse household energy behaviors. For the manufacturing use case, the FordA dataset supports binary classification to detect the presence of specific symptoms in an automotive subsystem.


\subsubsection{Implementation Setup}\label{section:implementation_setup}


The implementation setup includes the process of training AI models on the cloud server and testing on devices in the edge computing environment to collect the execution time traces.
In this study, we use an r6i.4xlarge EC2 instance on AWS to train and evaluate the accuracy of various privacy-preserving techniques across multiple AI models and datasets. Training is performed with a batch size of 50, over 25 epochs, and with a learning rate of 0.001. For a baseline comparison, we also measure the models' accuracy and inference times before applying any privacy techniques. Then, we conduct inference testing on an edge computing environment, comprising three edge devices and one edge server, as shown in Table \ref{tab:hardware_config}, to assess the real-world impact of privacy-preserving techniques in edge-based machine learning. The collected traces include 500 inference times from heterogeneous edge devices.

\begin{table}[h!]
  \centering
  \scriptsize
  \renewcommand{\arraystretch}{1.5} 
  \caption{Hardware Configuration}
  \label{tab:hardware_config}

  \begin{tabular}{|p{0.25\linewidth}|p{0.25\linewidth}|p{0.35\linewidth}|}
    \hline
    \textbf{Device} & \textbf{Type} & \textbf{Capacity} \\
    \hline
    EC2 Instance   & Cloud Server & 16 cores, 128 GB RAM, 3.7 GHz Xeon \\
    \hline
    Desktop        & Edge Server  & 16 cores, 32 GB RAM, 3.7 GHz Xeon \\
    \hline
    Intel NUC      & Edge Device  & 8 cores, 8 GB RAM, 1.1 GHz i7 \\
    \hline
    Jetson AGX     & Edge Device  & 8 cores, 32 GB RAM, ARMv8 \\
    \hline
    Raspberry Pi 5 & Edge Device  & 4 cores, 4 GB RAM, 2.4 GHz Cortex-A76 \\
    \hline
  \end{tabular}
\end{table}

\begin{table*}[t]
\centering
\scriptsize
\renewcommand{\arraystretch}{1.5} 
\caption{Implementation Parameters}
\begin{tabular}{|l|l|l|} \hline 
\textbf{Technique} & \textbf{Configuration Parameters} & \textbf{Tool}  \\ \hline  
\textbf{\gls{fhe}}        & \begin{tabular}[c]{@{}l@{}}Security bit (128), \\ Precision bit-width $p$ (5), \\ Quantization bit-width $q$ (4, 5, 6)\end{tabular}& Concrete-ML  \\ \hline  
\textbf{\gls{smc}}       & \begin{tabular}[c]{@{}l@{}} $world\_size$ (2, 3)\end{tabular} & Crypten      \\ \hline  
\textbf{\gls{dp}}        & \begin{tabular}[c]{@{}l@{}} Noise multiplier $\sigma$ (0.1, 0.2, 0.3, 0.4, 0.5, 0.6, 0.7),\\ Gradient clipping $C$ (1.0),\\ $\delta$ (1e-5)\end{tabular} & TensorFlow Privacy  \\ \hline 
\end{tabular}%
\label{tab:experiment_setup}
\end{table*}

Table \ref{tab:experiment_setup} provides a detailed configuration for each privacy-preserving technique, including the parameters for running experiments on various datasets. TensorFlow Privacy running \gls{dp} is configured to run on different noise multiplier levels, starting from 0.1 to 0.7. In the case of \gls{smc}, Crypten leverages multi-threading on a single EC2 instance, with two and three as the number of parties. Since the EC2 server experiment runs entirely on one EC2 instance, it focuses on computation cost and ignores communication overhead. This differs from a real-world edge computing setup, where each edge server or device would act as an independent party running its own instance of Crypten. In the hardware edge experimental setup, we evaluate the performance of \gls{smc} by applying deep learning models across two and three parties.

Concrete-ML enables \gls{ppml} with \gls{fhe} but is constrained by the maximum number of bits it can handle in its current library version \cite{ZamaConcreteML}. The current version of Concrete-ML 1.7 supports only a maximum of 8 bits. To meet this constraint and achieve practical performance, deep learning models must be quantized prior to execution. We explore different quantization levels in our setup, applying 4, 5, and 6 bits to reduce the model size while retaining accuracy. Additionally, the quantized model is compiled into an operational circuit graph that \gls{fhe} can interpret. During this compilation, we specify the number of precision bits used for privacy-preserving computation. 

\subsubsection{Simulation Setup}\label{AA}

To achieve a more realistic performance evaluation, we conducted trace-based simulations using real-world traces collected from the system implementation. EdgeSimPy \cite{souza2023edgesimpy}, a simulation toolkit for edge computing systems, is selected for our simulation approach. Developed in Python, the library also includes several built-in models for assessing power consumption and application composition for tracking the performance of edge computing applications.

Using the response time traces gathered from the hardware experiment described in Section \ref{section:implementation_setup}, we conducted different experiments in a simulated environment featuring four edge servers and twelve edge devices. The simulation framework, built using EdgeSimPy, was evaluated on three privacy-preserving techniques and executed on a desktop system equipped with a 16-core 3.7 GHz Xeon CPU and 128 GB of memory. Energy consumption was calculated based on the power usage of active devices during the simulation, which was monitored and recorded by EdgeSimPy. A linear server power model was employed to estimate power consumption, assuming a direct correlation between a server's power usage and workload demand. Additionally, a simple \textit{heuristic scheduler} was implemented, which allocates tasks to the nodes with the most available resources, prioritizing those with the highest combined CPU and memory availability. It is important to note that the scheduling algorithm was intentionally left unchanged to isolate and assess the performance impacts of \gls{ppml} techniques.

Table \ref{tab:simulation_config} provides a detailed overview of the configuration parameters, including the number of edge servers, devices, and the simulation setup.  
To introduce system heterogeneity, the four edge servers were equipped with 16 CPUs and 32GB of RAM each, while the 12 edge devices were randomly chosen from Intel NUCs, Jetson AGX boards, and Raspberry Pi 5s. Experiments simulated varying levels of concurrent user requests, starting at 30 and incrementing to 40, 50, 60, and 70, to evaluate performance under different workload intensities. These concurrent users affect the number of simultaneous requests received by the framework.
For the \gls{smc} technique, two network bandwidths (250 Mbps and 500 Mbps) were tested to assess the impact of \gls{smc} communication on the evaluation metrics. The simulation measured two key performance indicators: inference response times for each model and average energy consumption per inference, calculated using the maximum power draw ($P_{\text{max}}$) and idle power consumption ($P_{\text{static}}$) of each device as listed in Table \ref{tab:simulation_config}.

A comparative analysis is conducted across different privacy-preserving techniques and selected models against their baseline results measured without applying any privacy-preserving techniques. Table~\ref{tab:raw-results} shows the raw model performance, measured in terms of training time and accuracy on the EC2 instance, and inference time on the edge server. The results reveal that AlexNet generally provides the highest accuracy, especially for FordA, but with a higher inference time. LeNet-5 and SqueezeNet are more efficient in terms of inference time but may sacrifice some accuracy. Due to space constraints, we present only a subset of the results in the following sections; however, similar trends were observed in experiments conducted with other datasets.

\begin{table}[h]
    \caption{Simulation Parameters}
    \centering
    \scriptsize
    \renewcommand{\arraystretch}{1.5} 
    \begin{tabular}{|p{0.42\linewidth}|p{0.5\linewidth}|} \hline 
         \textbf{Parameter}& \textbf{Value}\\ \hline 
         Total Requests& 3000\\ \hline
         Response Time Traces& 500 data points\\ \hline
         Users& 30, 40, 50, 60, 70\\ \hline 
         Edge Nodes& 4 Edge Servers, 12 Edge Devices\\ \hline 
 Bandwidth (\gls{smc}) &250Mbps, 500Mbps\\\hline
 Parties (\gls{smc})& 2, 3 parties\\\hline
 Power Consumption &  Edge Server ($P_{\text{max}}$: 150W, $P_{\text{static}}$: 4.5W) \\
  &             Raspberry Pi 5 ($P_{\text{max}}$: 10W, $P_{\text{static}}$: 0.5W) \\
  &             NUC Intel ($P_{\text{max}}$: 32W, $P_{\text{static}}$: 6W)     \\ 
 &              Jetson AGX ($P_{\text{max}}$: 11.6W, $P_{\text{static}}$: 2.75W) \\ \hline 
\gls{fhe} bit widths & 4-bit quantization; 5-bit precision \\ \hline
 
    \end{tabular}
    \vspace{10pt}  
    \label{tab:simulation_config}
\end{table}

\section{Results and Discussion}
\label{section:results}

This section presents real-system experimentation results (edge devices and servers), then trace-based simulation results using the collected traces, and finally a detailed discussion of the evaluation.

\subsection{Empirical Evaluation on Edge Platforms} 

This section reports empirical results for three privacy-preserving techniques on edge platforms. 
First, we quantify the accuracy impact of \gls{dp} and \gls{fhe} across models and datasets, varying noise levels for \gls{dp} and quantization bit-widths for \gls{fhe}, measured on edge servers. 
Next, we examine \gls{fhe} accuracy–latency trade-offs under different quantization and numeric precision settings, also on edge servers. 
Finally, we study how party size impacts \gls{smc} inference time across models and datasets, with one edge server as the first party and additional edge servers as the remaining parties.

\begin{figure}[!htbp]
  \centering

  \begin{subfigure}[t]{0.75\columnwidth}
    \centering
    \includegraphics[width=\linewidth]{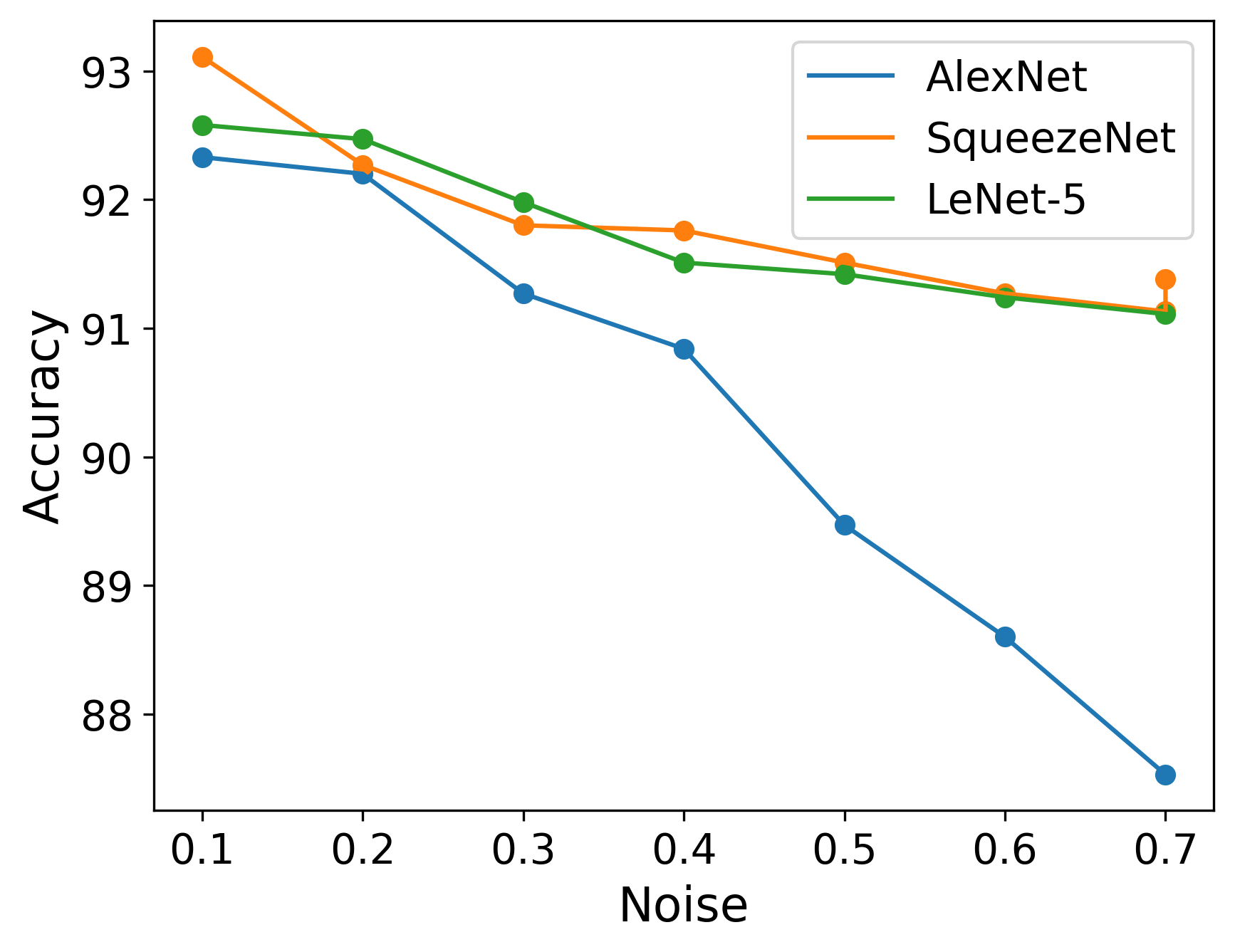}
    \caption{\gls{dp} accuracy (ECG5000)}
    \label{fig:acc-dp-a}
  \end{subfigure}\hfill
  \begin{subfigure}[t]{0.75\columnwidth}
    \centering
    \includegraphics[width=\linewidth]{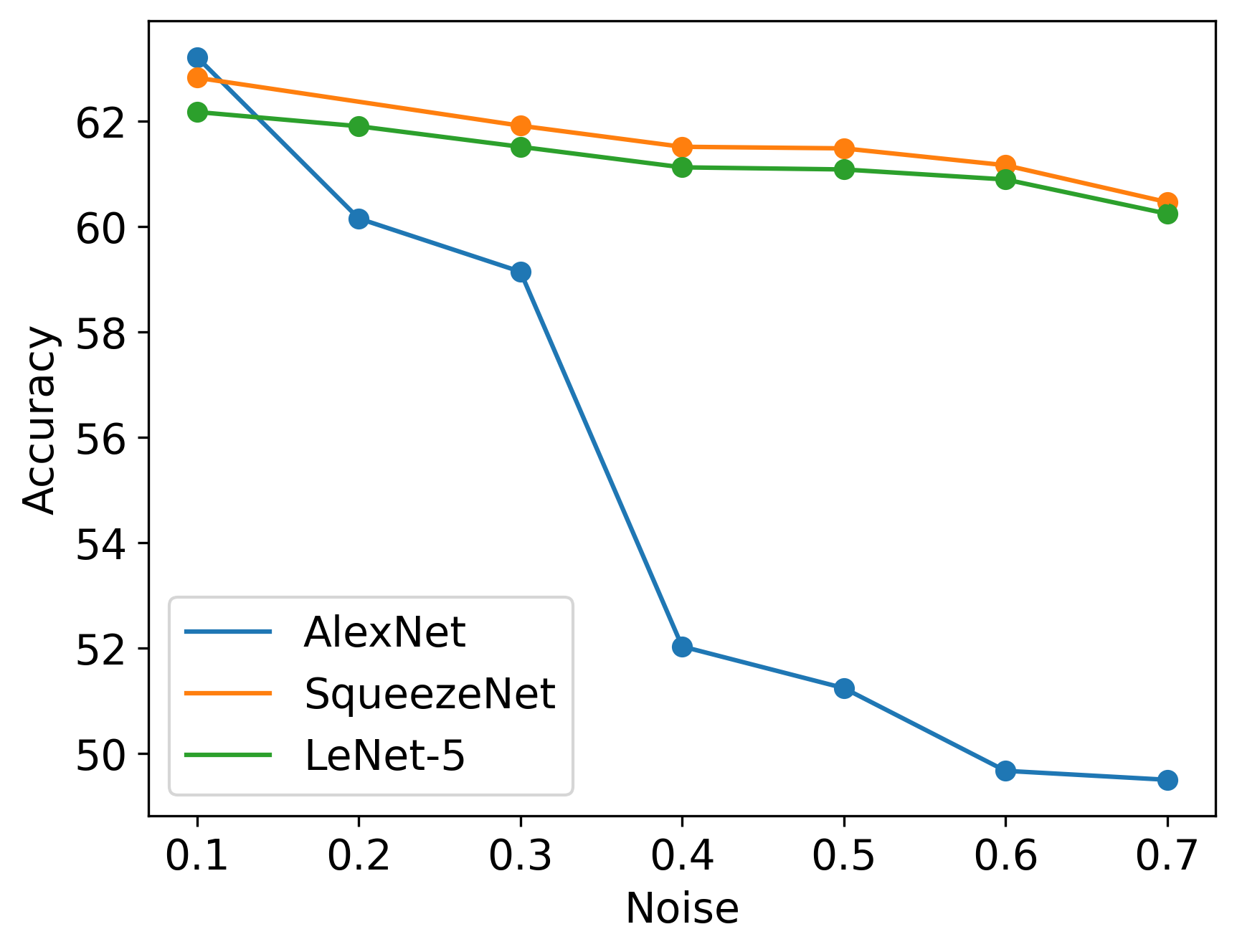}
    \caption{\gls{dp} accuracy (ElectricDevices)}
    \label{fig:acc-dp-b}
  \end{subfigure}

  \vspace{0.4em}

  \begin{subfigure}[t]{0.75\columnwidth}
    \centering
    \includegraphics[width=\linewidth]{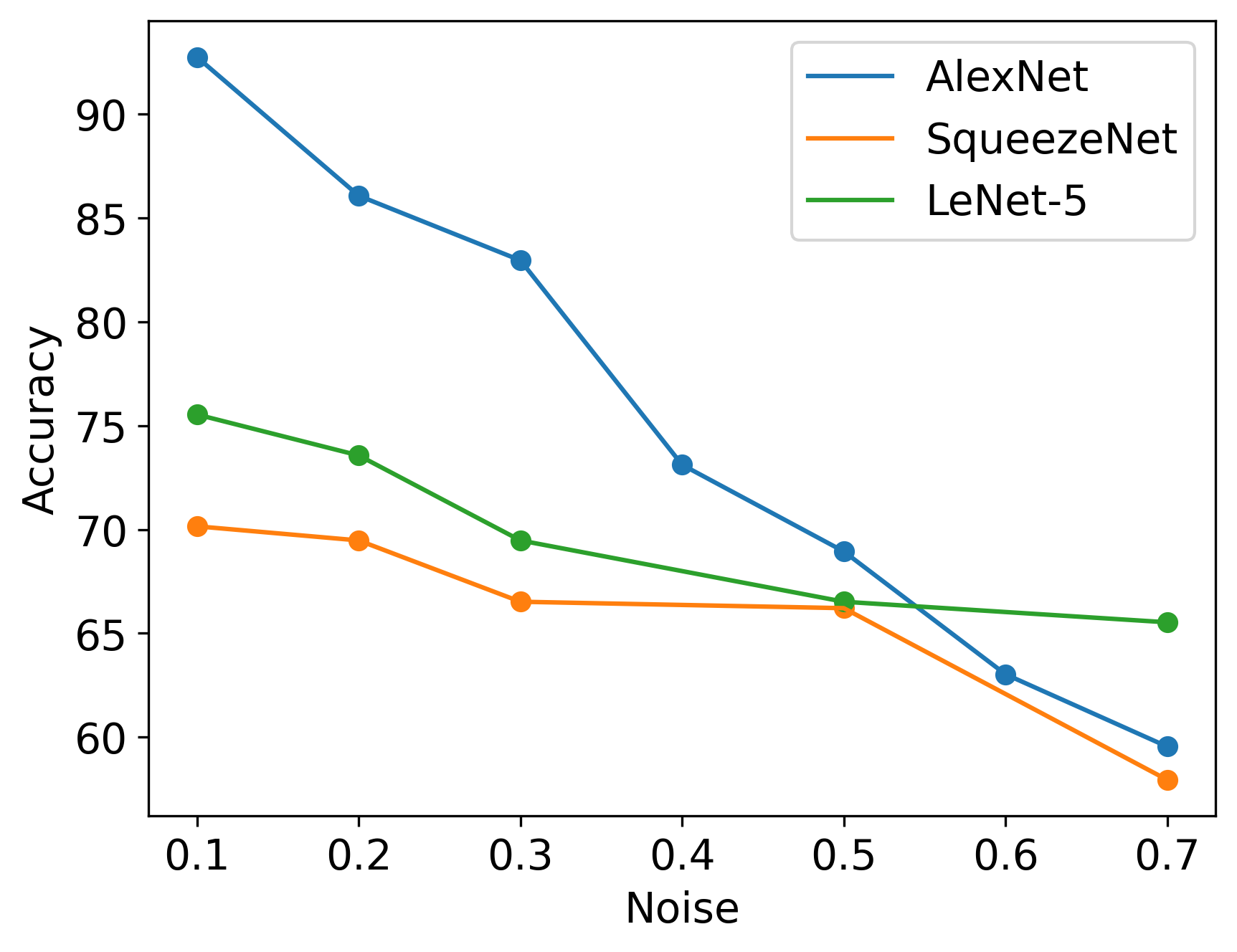}
    \caption{\gls{dp} accuracy (FordA)}
    \label{fig:acc-dp-c}
  \end{subfigure}\hfill
  \begin{subfigure}[t]{0.75\columnwidth}
    \centering
    \includegraphics[width=\linewidth]{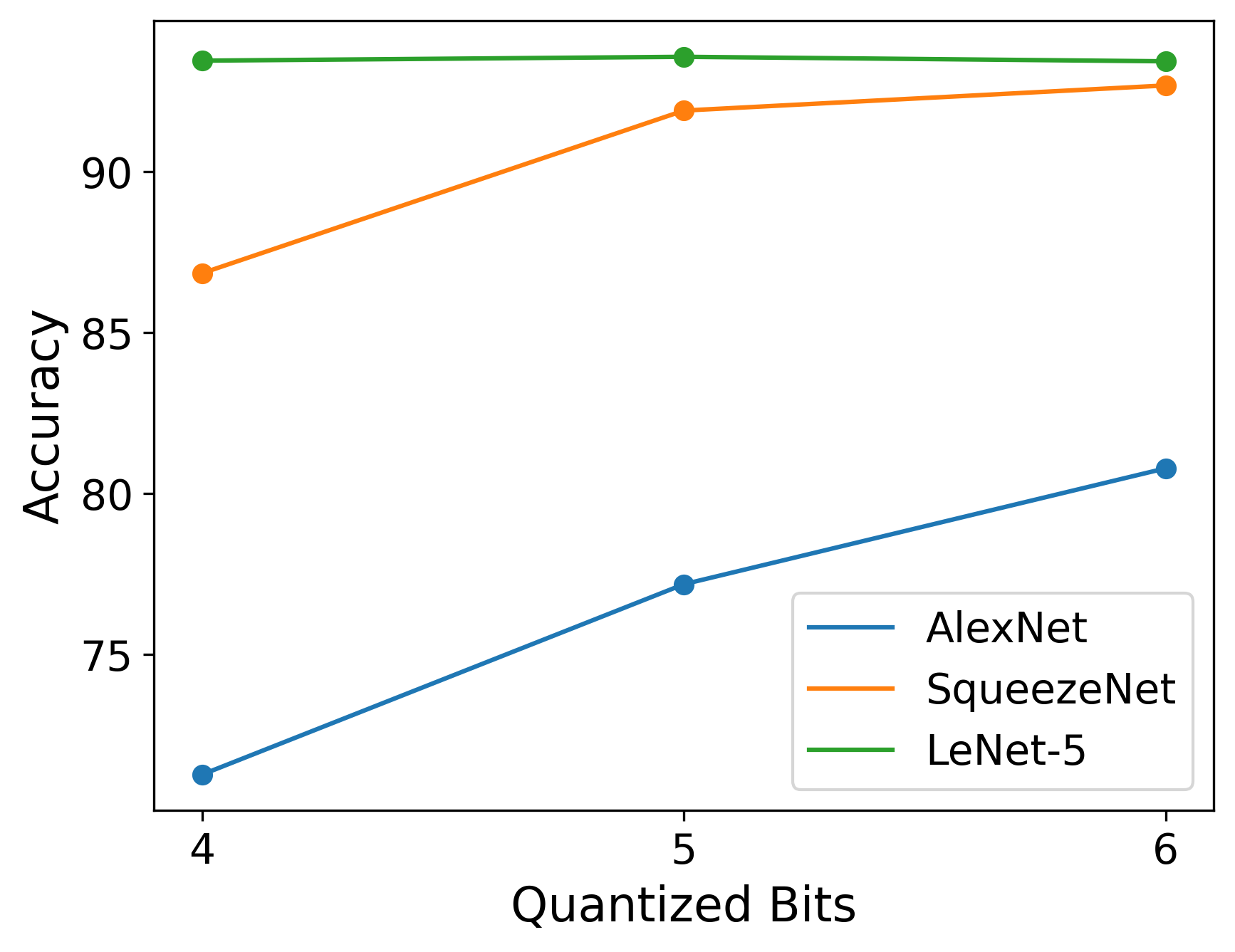}
    \caption{\gls{fhe} accuracy (ECG5000)}
    \label{fig:acc-fhe-ecg}
  \end{subfigure}

  \caption{Accuracy of privacy-preserving techniques across datasets (\gls{dp} and \gls{fhe}).}
  \label{fig:accuracy-ecg5000}
\end{figure}

The \gls{dp} technique enhances privacy by adding noise to the gradients during training, but this comes at the cost of accuracy as shown in Figs.~\ref{fig:acc-dp-a}--\ref{fig:acc-dp-c}.
Accuracy declines monotonically with the noise multiplier, with sensitivity to both model and dataset.
AlexNet is the most sensitive: on ECG5000 its accuracy drops by 3--4\% as noise increases from 0.1 to 0.7, by 21--22\% on ElectricDevices, and by 35\% on FordA.
By contrast, LeNet-5 and SqueezeNet are more robust: 1--2\% on ECG5000, $\le$3\% on ElectricDevices, and 13--18\% on FordA.
This argues that we should set the \gls{dp} budget separately for each model and dataset, not use one value for everything.

For the \gls{fhe} technique in Concrete-ML with 5-bit precision (Figure \ref{fig:acc-fhe-ecg}), accuracy improves with quantization bit width and shows diminishing returns beyond 5 bits.
From 4 to 6 bits, AlexNet and SqueezeNet gain roughly 3--7\%, while LeNet-5 changes by $\le$1\%. At 6 bits, all models are close to their non-encrypted accuracy (In Table~\ref{tab:raw-results}), at the cost of higher compute, since \gls{fhe} latency generally grows with precision.

DP can induce sizable drops on some (model,\,dataset) pairs, most notably AlexNet on FordA and ElectricDevices, while FHE with 5--6 bits typically retains accuracy within a few percent of baseline on ECG5000.
Thus, when accuracy headroom is tight, favor \gls{fhe} (or \gls{smc}); when inference cost must stay minimal, \gls{dp} is attractive but requires careful noise calibration.
It is worth noting that \gls{smc} is omitted here because its inference accuracy is essentially indistinguishable.


\begin{table*}[t]
\centering
\scriptsize
\renewcommand{\arraystretch}{1.5} 
\caption{Raw Model Performance on Various Datasets}
\begin{tabular}{|l|l|c|c|c|}
\hline
\textbf{Dataset} & \textbf{Model} & \textbf{Accuracy (\%)} & \textbf{Inference (ms)} & \textbf{Training (s)} \\ \hline
\multirow{3}{*}{ECG5000}         
    & LeNet-5  & 93.13  & 45.9  & 14.0  \\ 
    & SqueezeNet    & 93.53  & 46.8  & 109.5 \\ 
    & AlexNet  & 95.80  & 58.5  & 169.6 \\ \hline
\multirow{3}{*}{ElectricDevices}  
    & LeNet-5  & 67.00  & 44.2  & 52.8  \\ 
    & SqueezeNet    & 65.66  & 44.9  & 293.2 \\ 
    & AlexNet  & 64.44  & 56.0  & 631.2 \\ \hline
\multirow{3}{*}{FordA}             
    & LeNet-5  & 80.38  & 46.7  & 56.1  \\ 
    & SqueezeNet    & 77.50  & 48.6  & 388.7 \\ 
    & AlexNet  & 92.31  & 64.0  & 499.4 \\ \hline
\end{tabular}%
\label{tab:raw-results}
\end{table*}

\begin{figure}[h]
  \centering
  \begin{subfigure}[t]{\linewidth}
    \centering
    \includegraphics[width=0.9\linewidth]{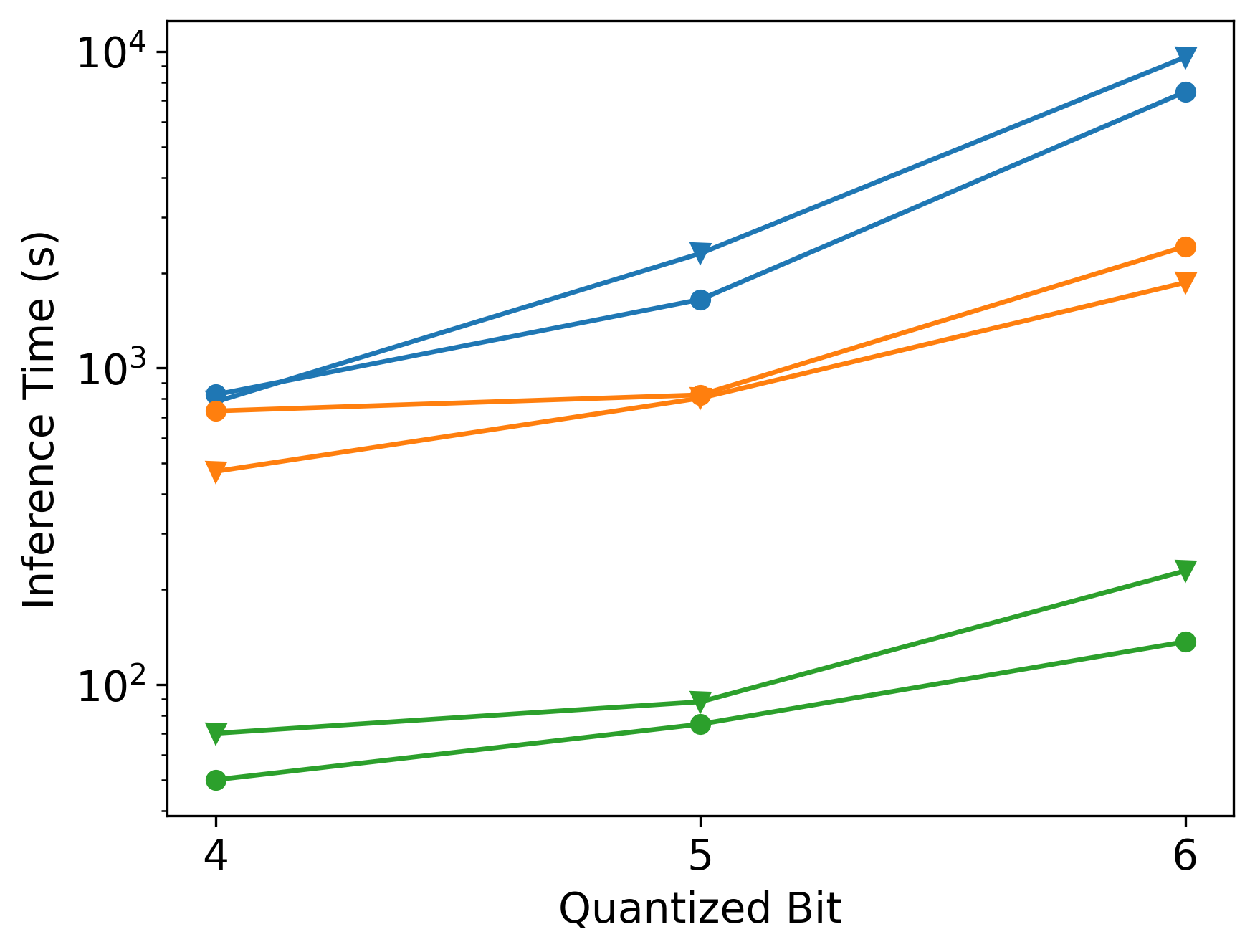}
    \caption{Inference Time (ElectricDevices)}
    \label{fig:fhe_bit_widths_electric}
  \end{subfigure}
  \hfill
\begin{subfigure}[t]{\linewidth}
    \centering
    \includegraphics[width=0.9\linewidth]{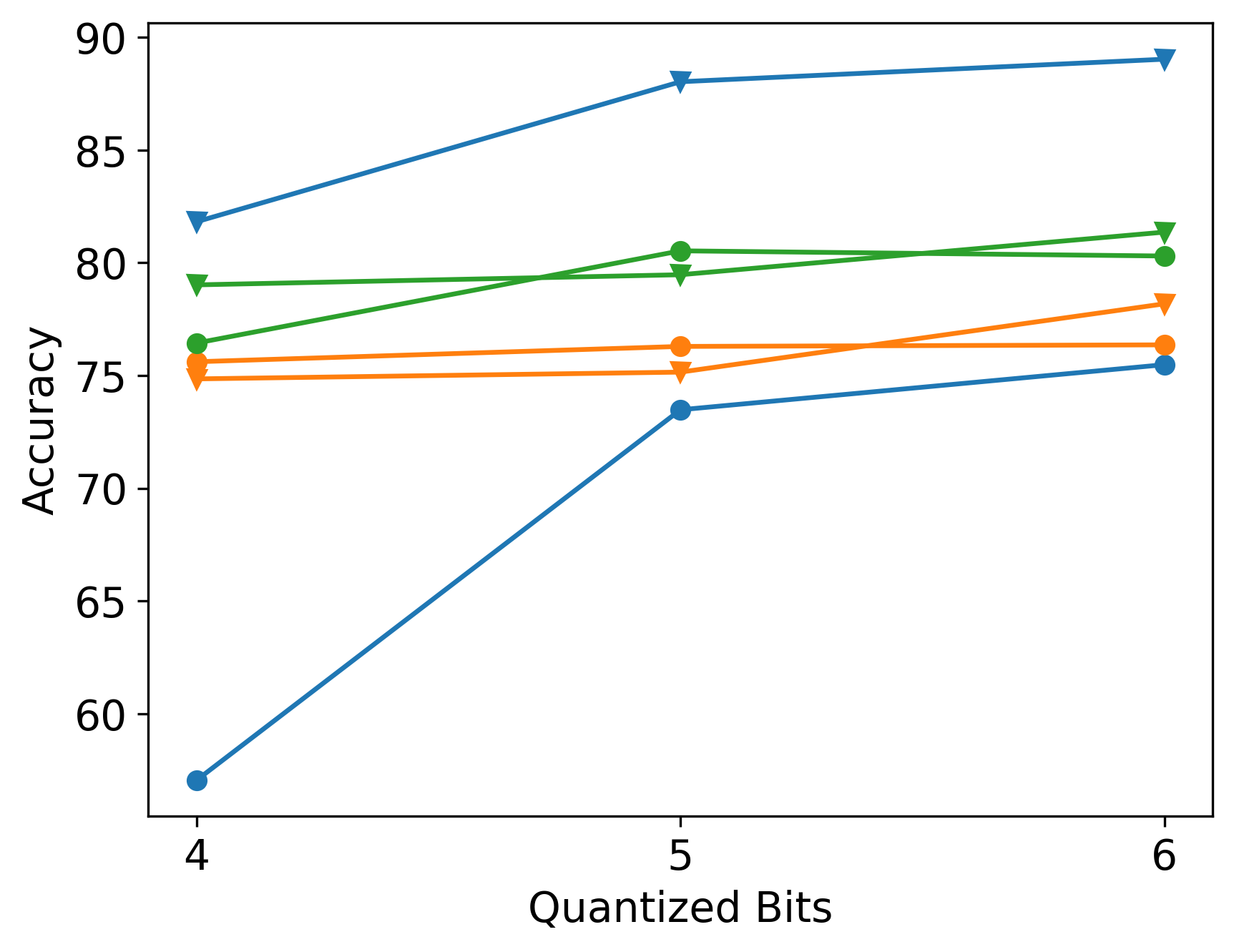}
    \caption{Accuracy (FordA)}
    \label{fig:fhe_bit_widths_forda}
  \end{subfigure}

  \begin{subfigure}[t]{\linewidth}
    \centering
    \includegraphics[width=\linewidth]{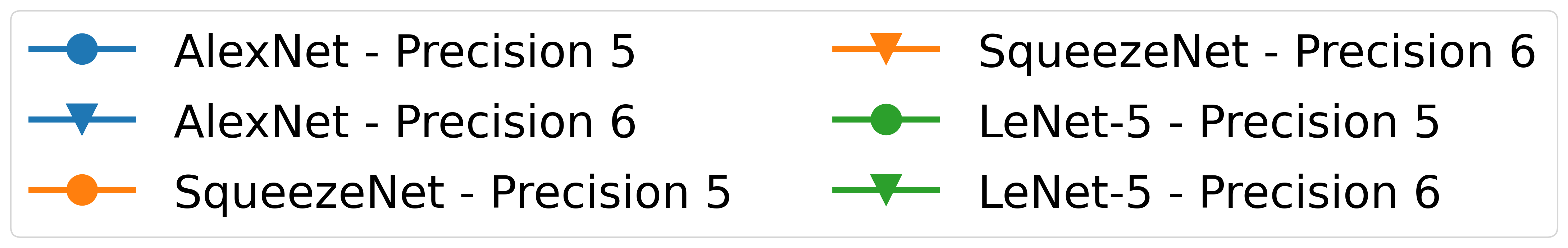}
  \end{subfigure}

  \caption{Effect of precision and quantization bit widths on \gls{fhe} inference time and accuracy across datasets.}
  \label{fig:fhe_bit_widths}
\end{figure}

Figure~\ref{fig:fhe_bit_widths} isolates the accuracy–latency trade-off of \gls{fhe} by varying quantization bit width (4--6) and numeric precision (5 vs.\ 6). The left panel (\ref{fig:fhe_bit_widths_electric}) plots inference time on a log scale for ElectricDevices; the right (\ref{fig:fhe_bit_widths_forda}) shows accuracy on FordA.

Regarding the latency scaling (Fig.~\ref{fig:fhe_bit_widths_electric}),
Inference time grows super linearly with quantization bit width, and the growth factor depends strongly on the model:
AlexNet increases by roughly 7--9 times from 4 to 6 bits,
SqueezeNet by 3--4 times,
and LeNet-5 by 2--3 times.
At any fixed bit width, moving from precision~5 to~6 further increases runtime (typically 20--60\%, and largest for AlexNet),
so precision amplifies the cost of higher bit widths. Model complexity is the dominant factor: AlexNet $\gg$ SqueezeNet $>$ LeNet-5 in absolute latency.

Figure~\ref{fig:fhe_bit_widths_forda} shows accuracy generally rising with quantized bit-width: most gains occur from 4 to 5 bits, with smaller changes from 5 to 6.
Across 4 to 6 bits, AlexNet improves by \(10\text{--}20\%\), SqueezeNet by \(4\text{--}9\%\), and LeNet-5 by \(\le 5\%\).
Changing numeric precision (5–6) has only minor, model-dependent effects for SqueezeNet and LeNet-5, but a noticeable impact for AlexNet, likely due to its more complex architecture. For AlexNet, precision~6 trails precision~5 by \( 8\text{--}15\%\) at each bit-width; for SqueezeNet and LeNet-5, precision~6 is slightly higher by \( 1\text{--}2\%\).

Bit width drives most accuracy gains but incurs superlinear latency; precision~6 adds further cost. If latency-bound, choose the smallest bit width meeting the target (often 5) and lighter models. LeNet-5 is near-saturated at 4--5 bits and much faster, making it a strong FHE choice for constrained edge platforms.

\begin{figure*}[t]
  \centering

  \begin{subfigure}[t]{0.30\linewidth}
    \centering
    \includegraphics[width=\linewidth]{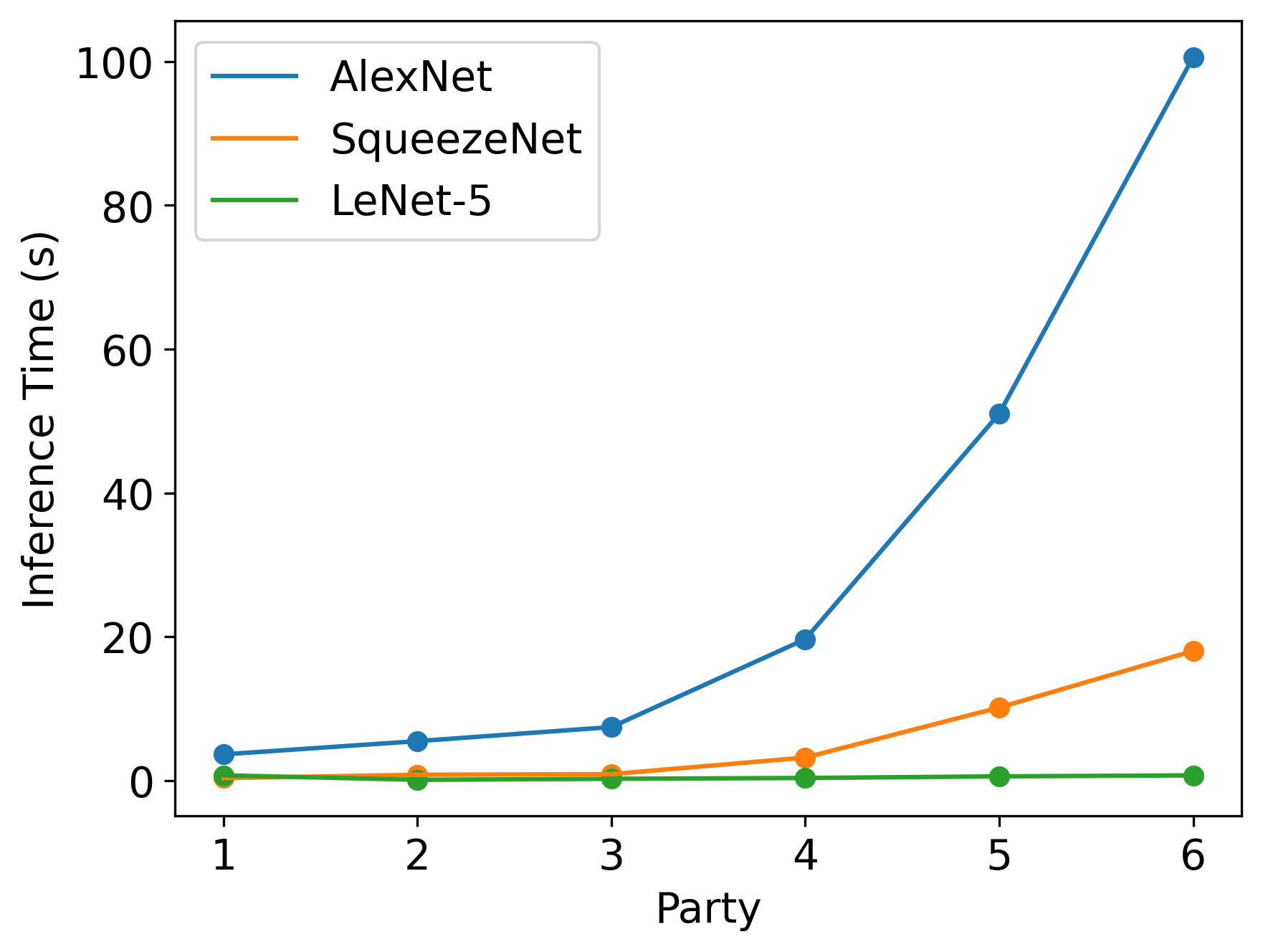}
    \caption{ECG5000}
    \label{fig:smc_ecg5000}
  \end{subfigure}\hfill
  \begin{subfigure}[t]{0.30\linewidth}
    \centering
    \includegraphics[width=\linewidth]{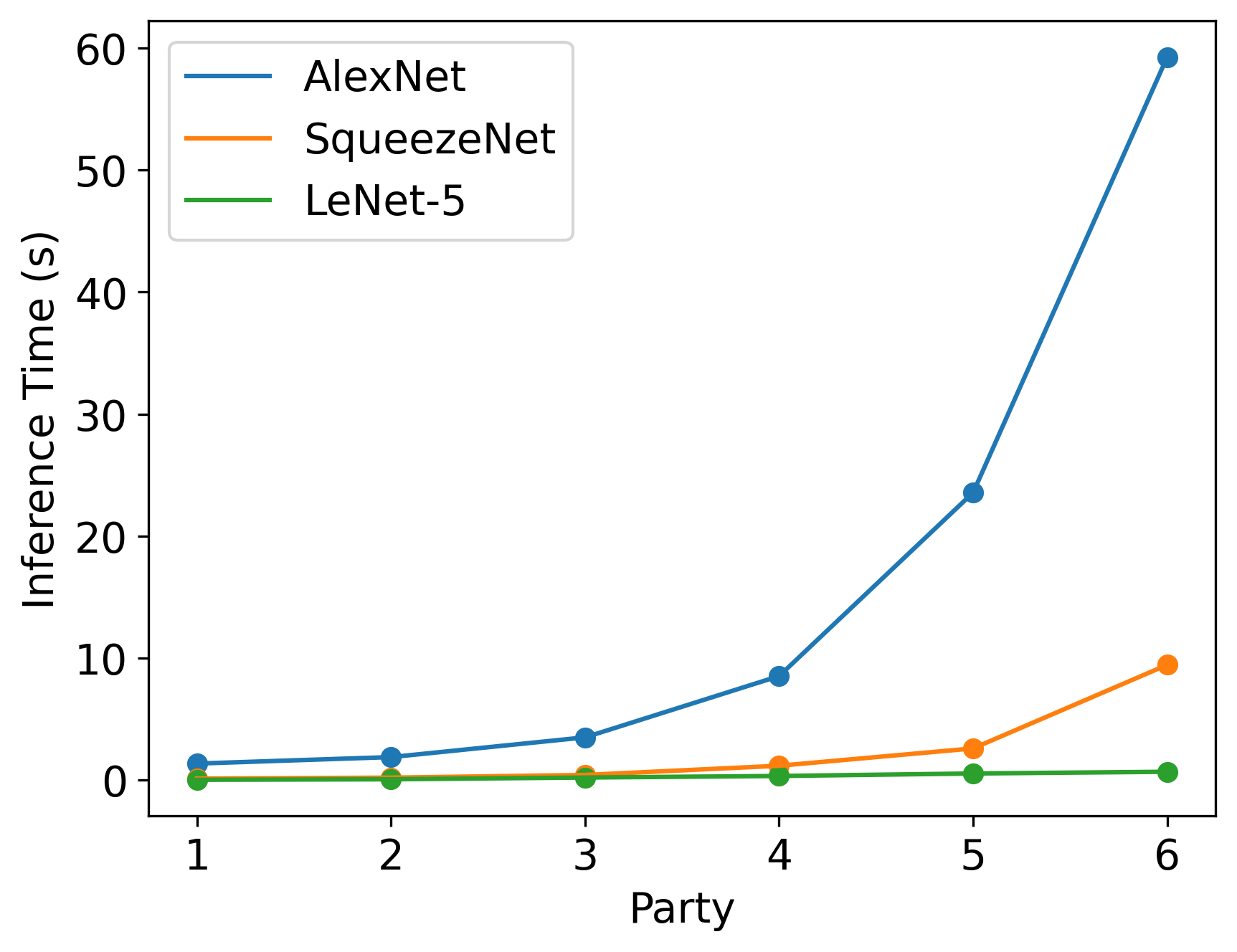}
    \caption{ElectricDevices}
    \label{fig:smc_electric}
  \end{subfigure}\hfill
  \begin{subfigure}[t]{0.30\linewidth}
    \centering
    \includegraphics[width=\linewidth]{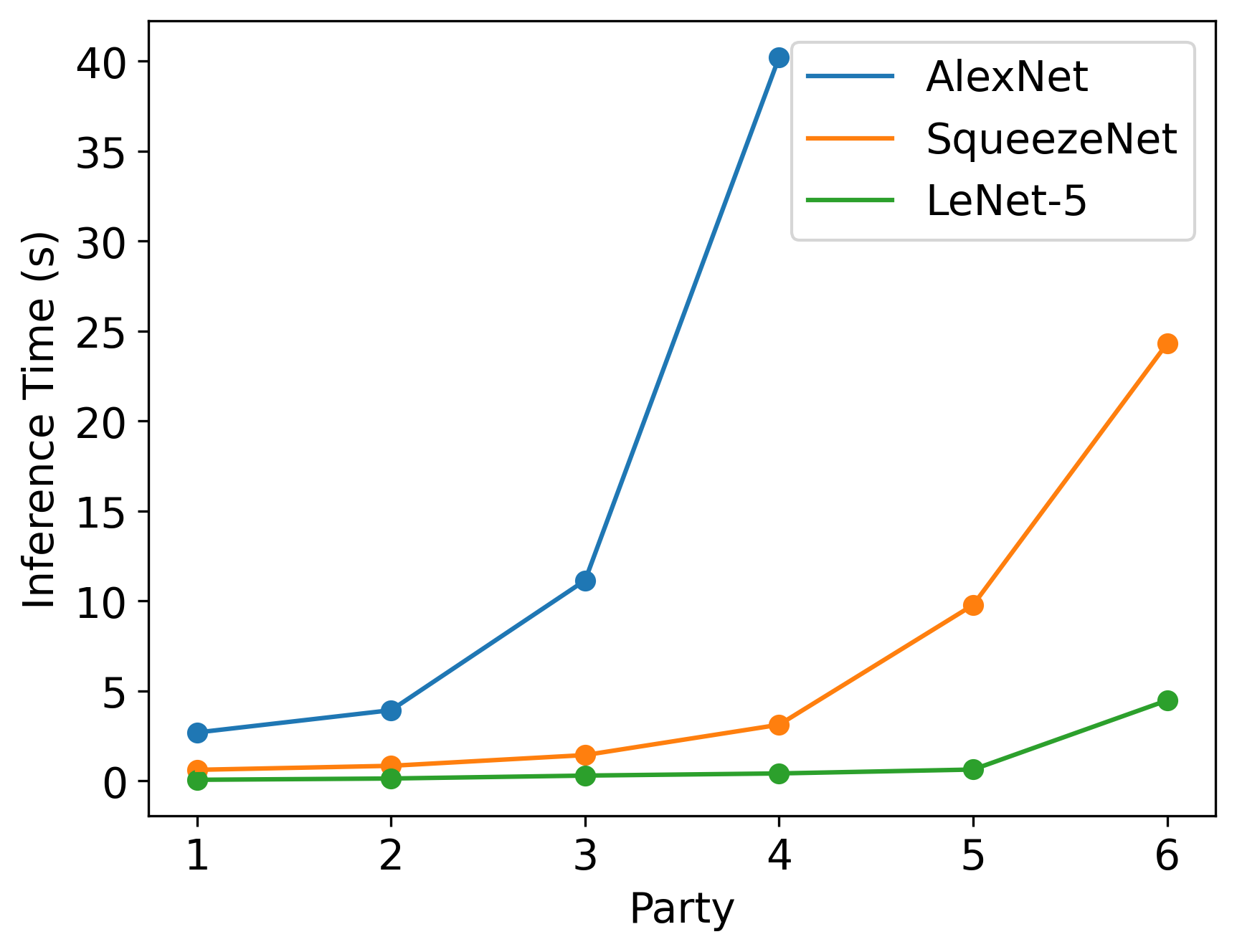}
    \caption{FordA}
    \label{fig:smc_forda}
  \end{subfigure}

  \caption{Effect of parties on \gls{smc} inference time across models and datasets.}
  \label{fig:smc_parties}
\end{figure*}

We also evaluate how party size affects \gls{smc} latency. As shown in Fig.~\ref{fig:smc_parties}, inference time grows at least linearly with the number of parties due to added interactive multiplications and communication rounds. The effect is architecture- rather than dataset-driven: AlexNet accelerates sharply beyond 4 parties, SqueezeNet grows moderately, and LeNet-5 remains nearly flat. In practice, choose the smallest size that meets the threat model and prefer lighter architectures or reduced multiplicative depth to limit rounds and traffic.

\subsection{Performance Analysis} 

\begin{figure*}[h]
  \centering
  \begin{subfigure}[t]{0.32\linewidth}
    \centering
    \includegraphics[width=\linewidth]{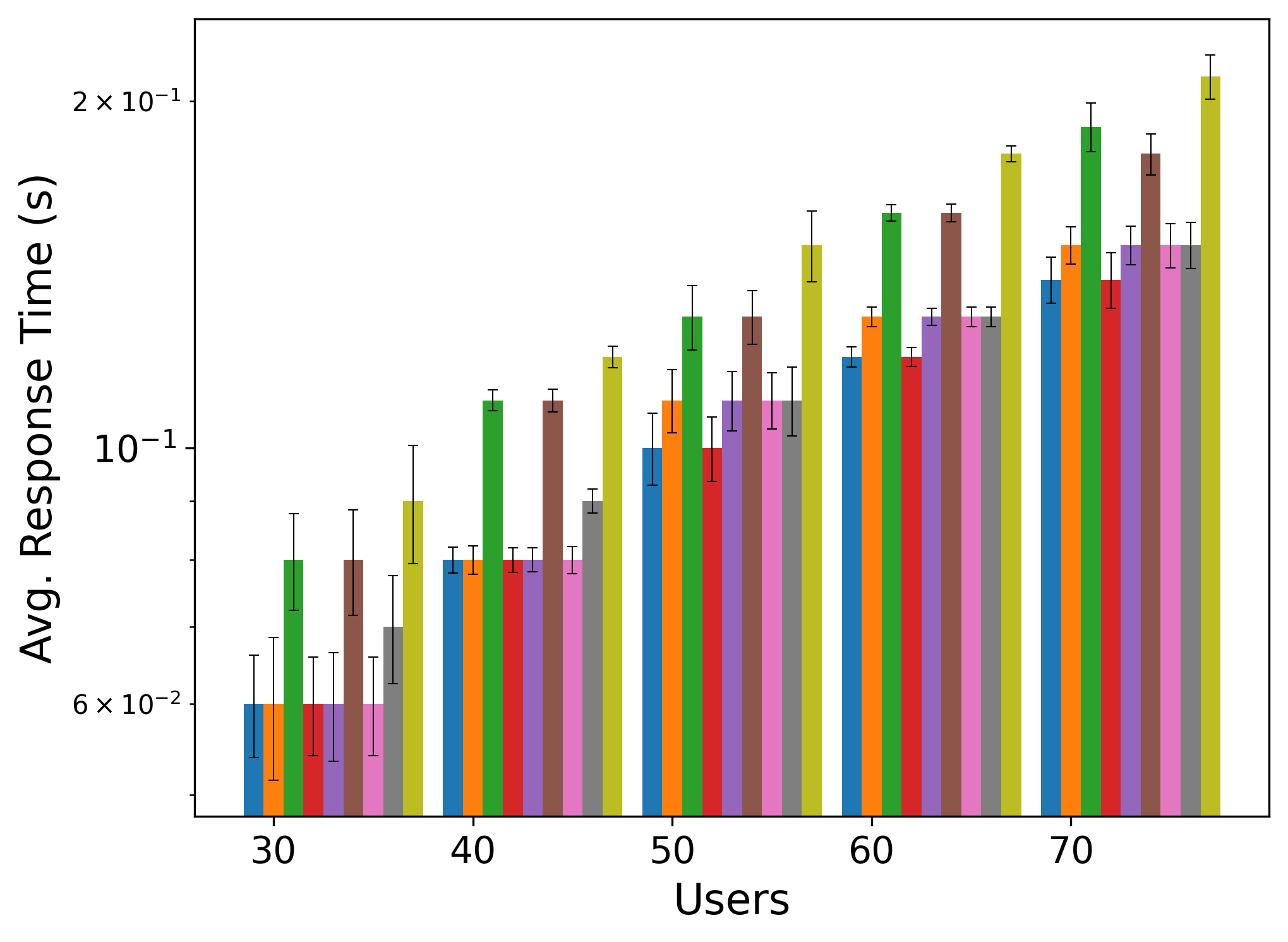}
    \caption{\gls{dp} average responses for different concurrent users}
    \label{fig:dp-avgres}
  \end{subfigure} \hfill
  \begin{subfigure}[t]{0.32\linewidth}
    \centering
    \includegraphics[width=\linewidth]{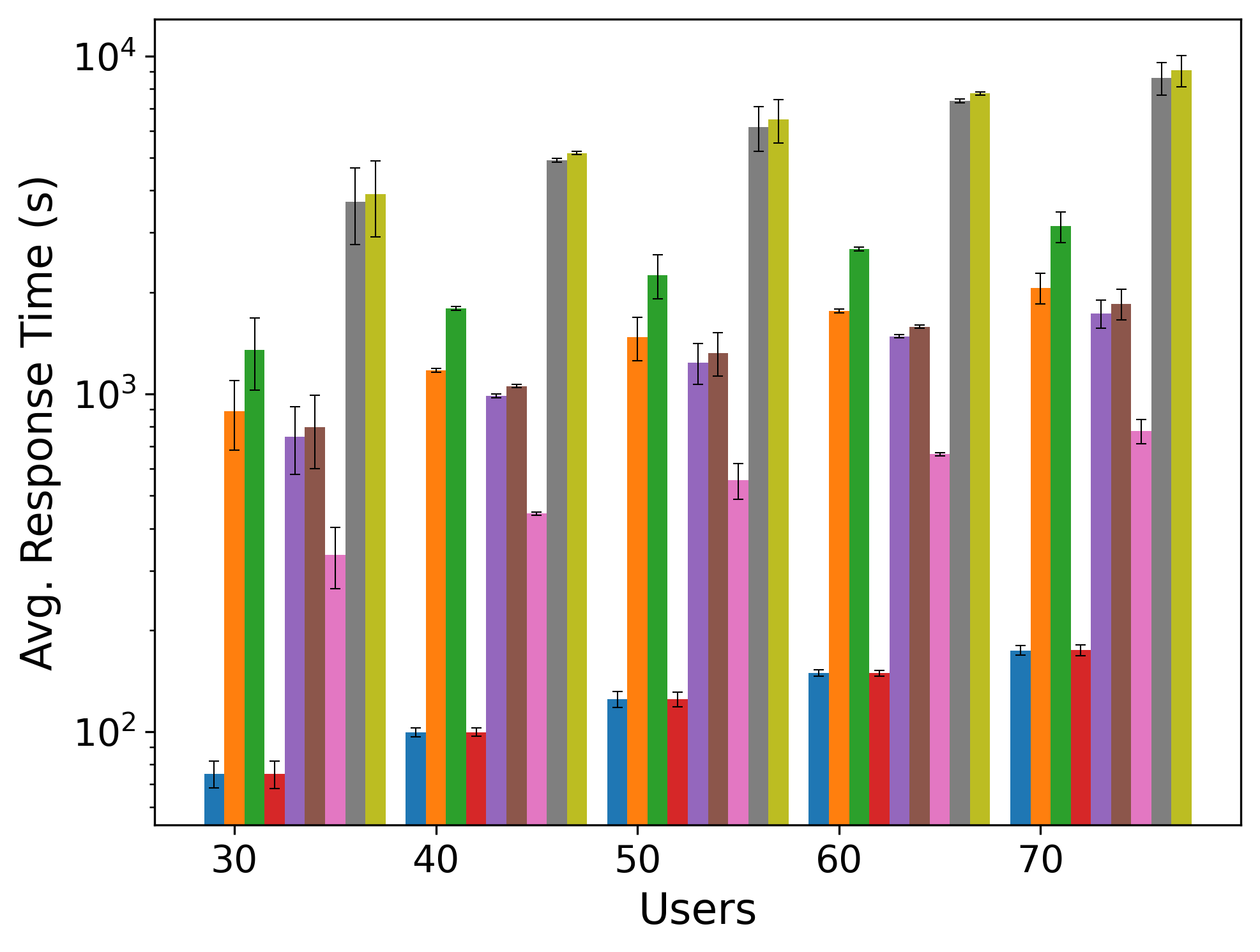}
    \caption{\gls{fhe} average responses for different concurrent users}
    \label{fig:concrete-avgres}
  \end{subfigure} \hfill
  \begin{subfigure}[t]{0.32\linewidth}
    \centering
    \includegraphics[width=\linewidth]{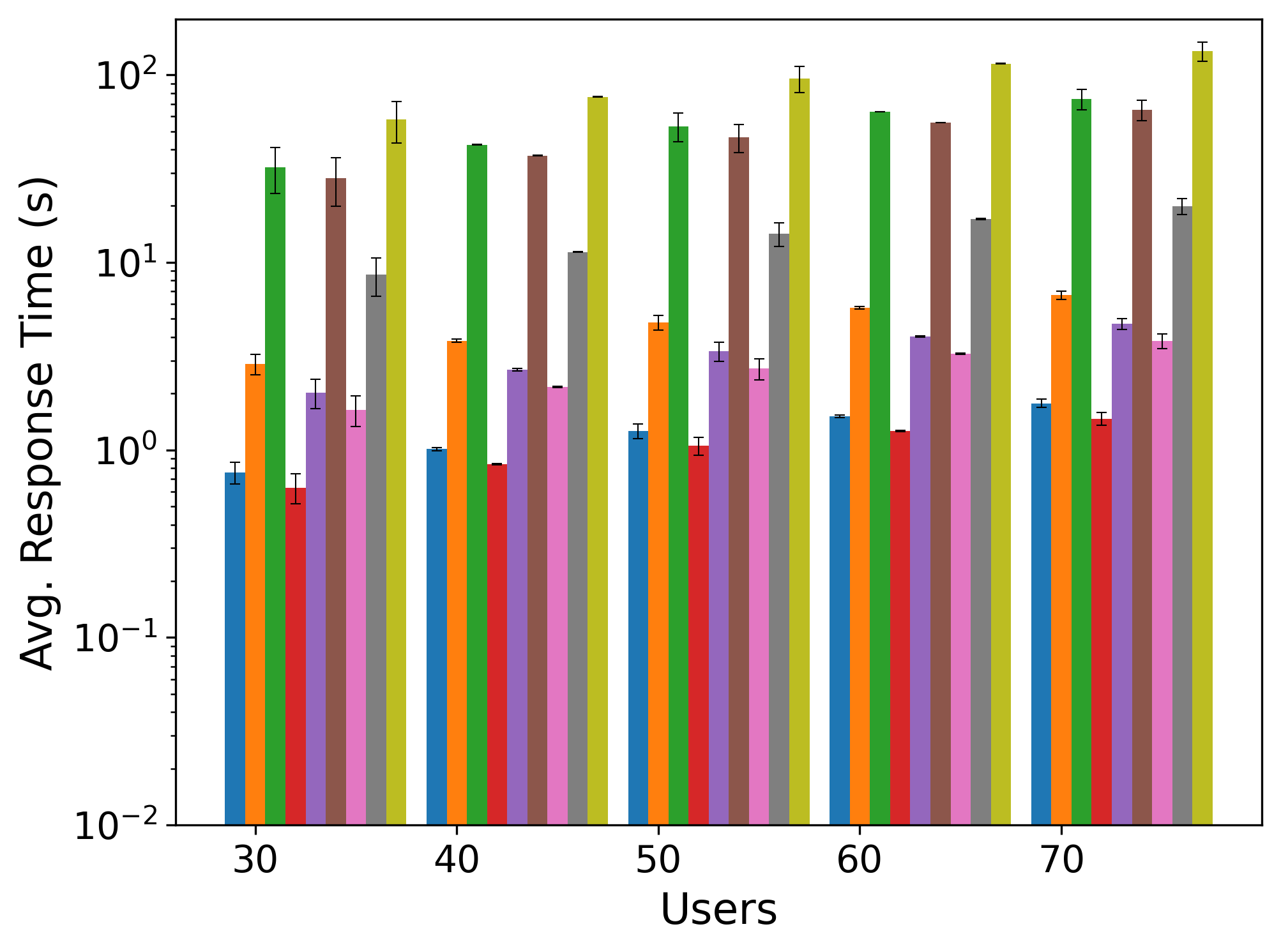}
    \caption{\gls{smc} two-party average responses in 250Mbps bandwidth}
    \label{fig:subfig-two-party-250}
  \end{subfigure}

  \vspace{1pt} 

  \begin{subfigure}[t]{0.32\linewidth}
    \centering
    \includegraphics[width=\linewidth]{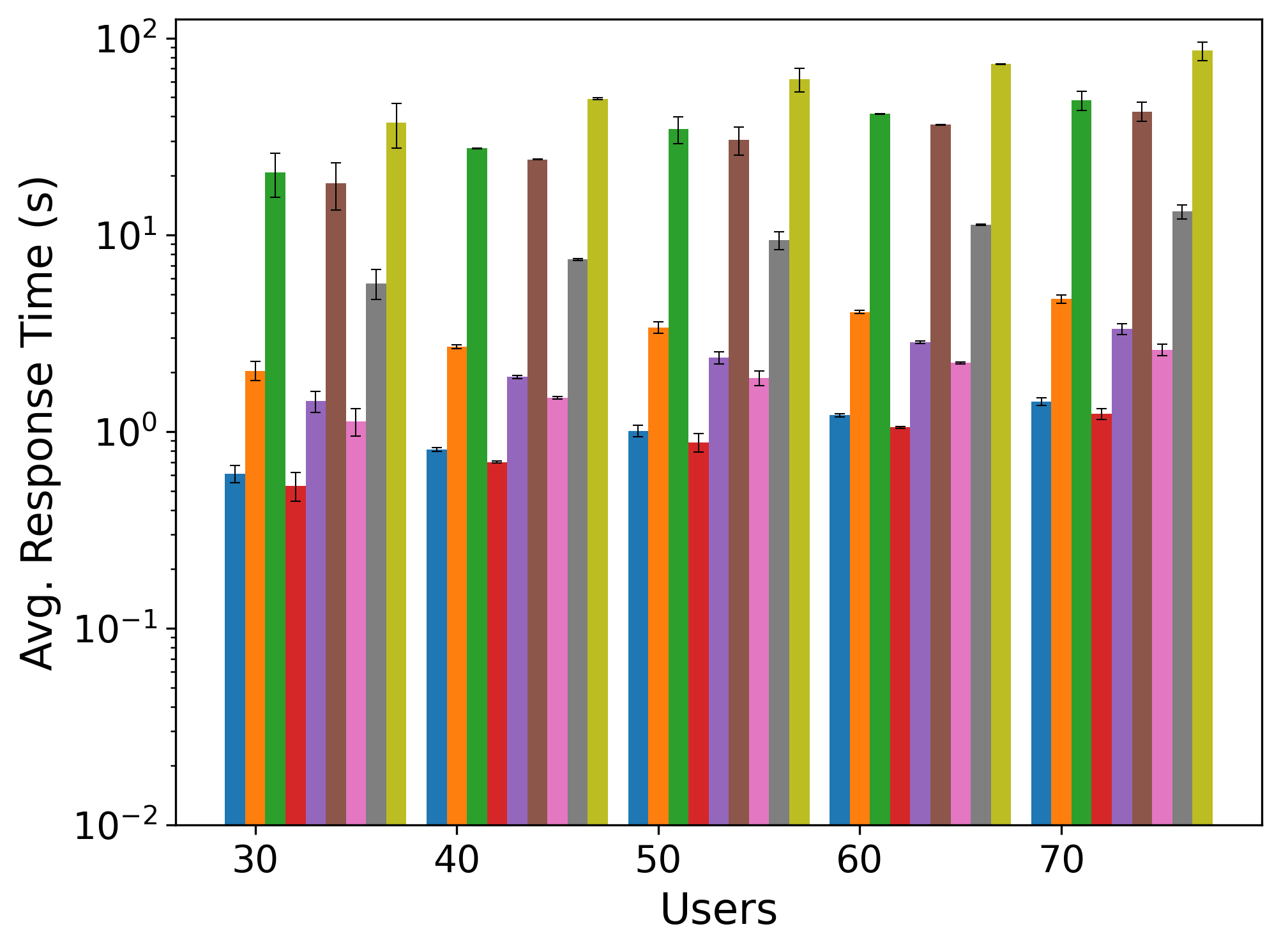}
    \caption{\gls{smc} two-party average responses in 500Mbps bandwidth}
    \label{fig:subfig-two-party-500}
  \end{subfigure} \hfill
  \begin{subfigure}[t]{0.32\linewidth}
    \centering
    \includegraphics[width=\linewidth]{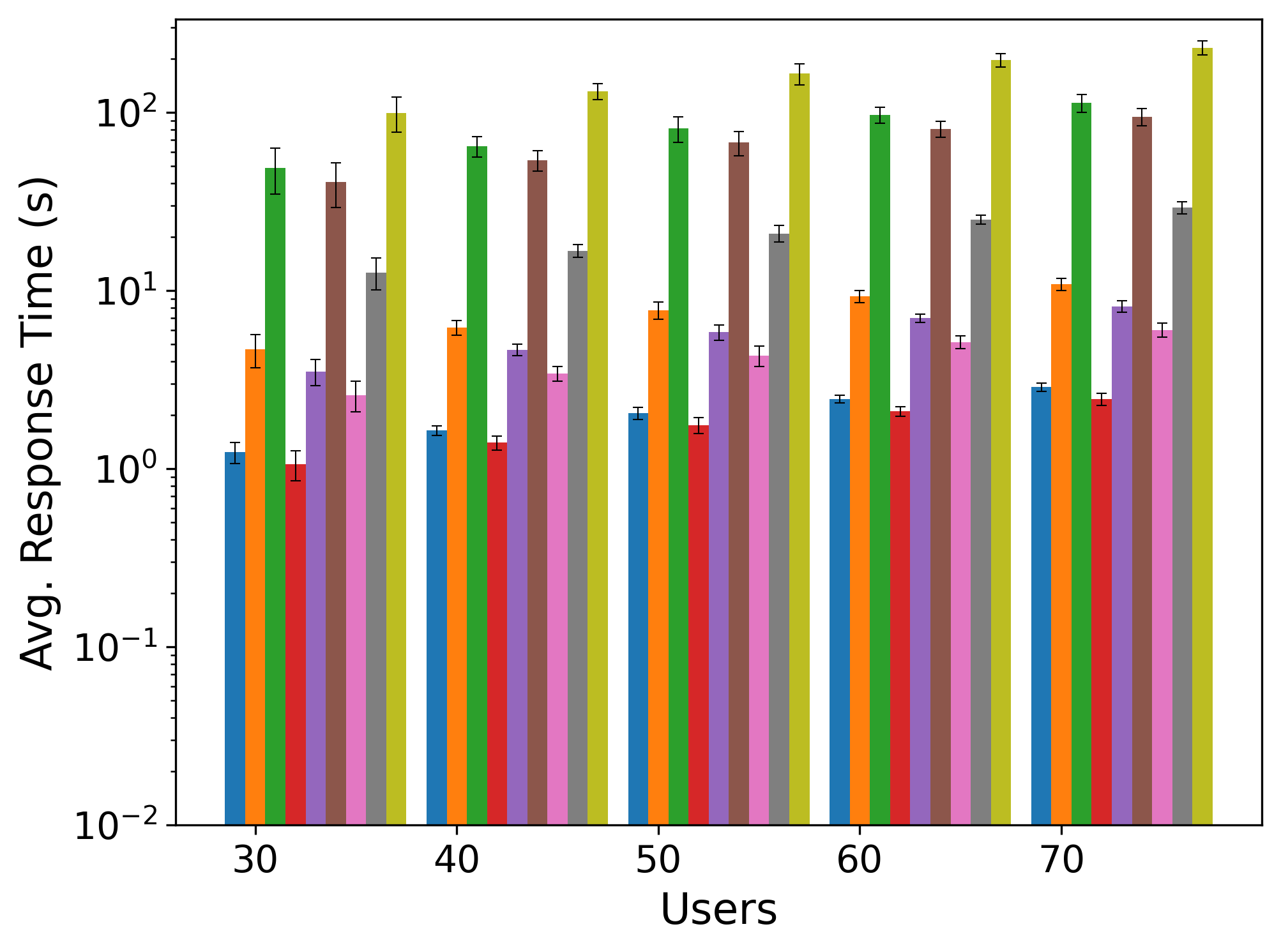}
    \caption{\gls{smc} three-party average responses in 250Mbps bandwidth}
    \label{fig:subfig-three-party-250}
  \end{subfigure} \hfill
  \begin{subfigure}[t]{0.32\linewidth}
    \centering
    \includegraphics[width=\linewidth]{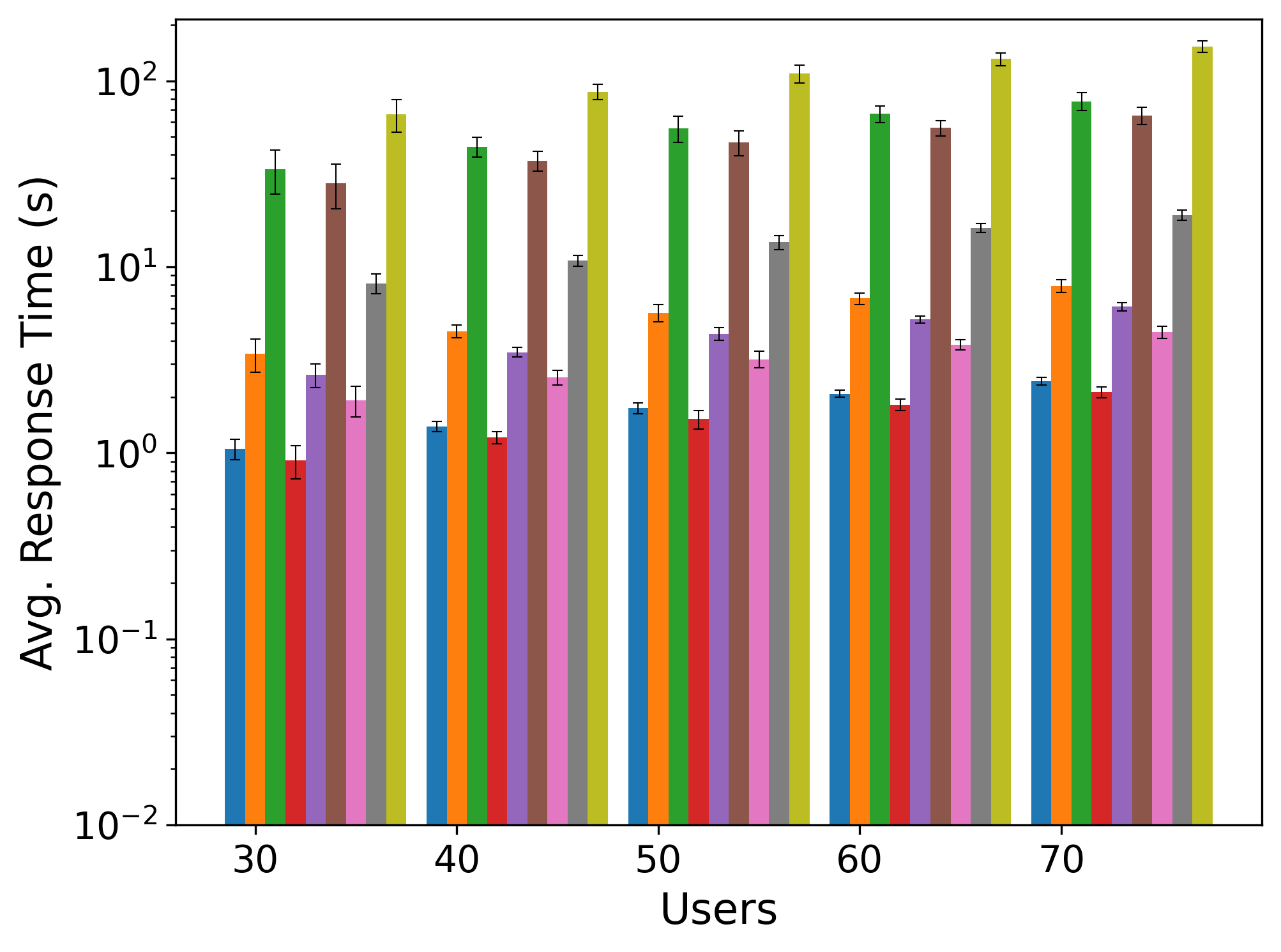}
    \caption{\gls{smc} three-party average responses in 500Mbps bandwidth}
    \label{fig:subfig-three-party-500}
  \end{subfigure}
  
  \vspace{1pt}

  \begin{subfigure}[t]{\linewidth}
    \centering
    \includegraphics[width=\linewidth]{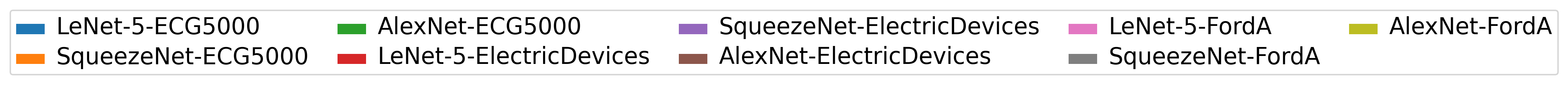}
  \end{subfigure}

  \caption{Average response times of privacy-preserving techniques with different numbers of concurrent users}
  \label{fig:avg-performance}
\end{figure*}

\begin{figure*}[h]
  \centering
  \begin{subfigure}[t]{0.32\linewidth}
    \centering
    \includegraphics[width=\linewidth]{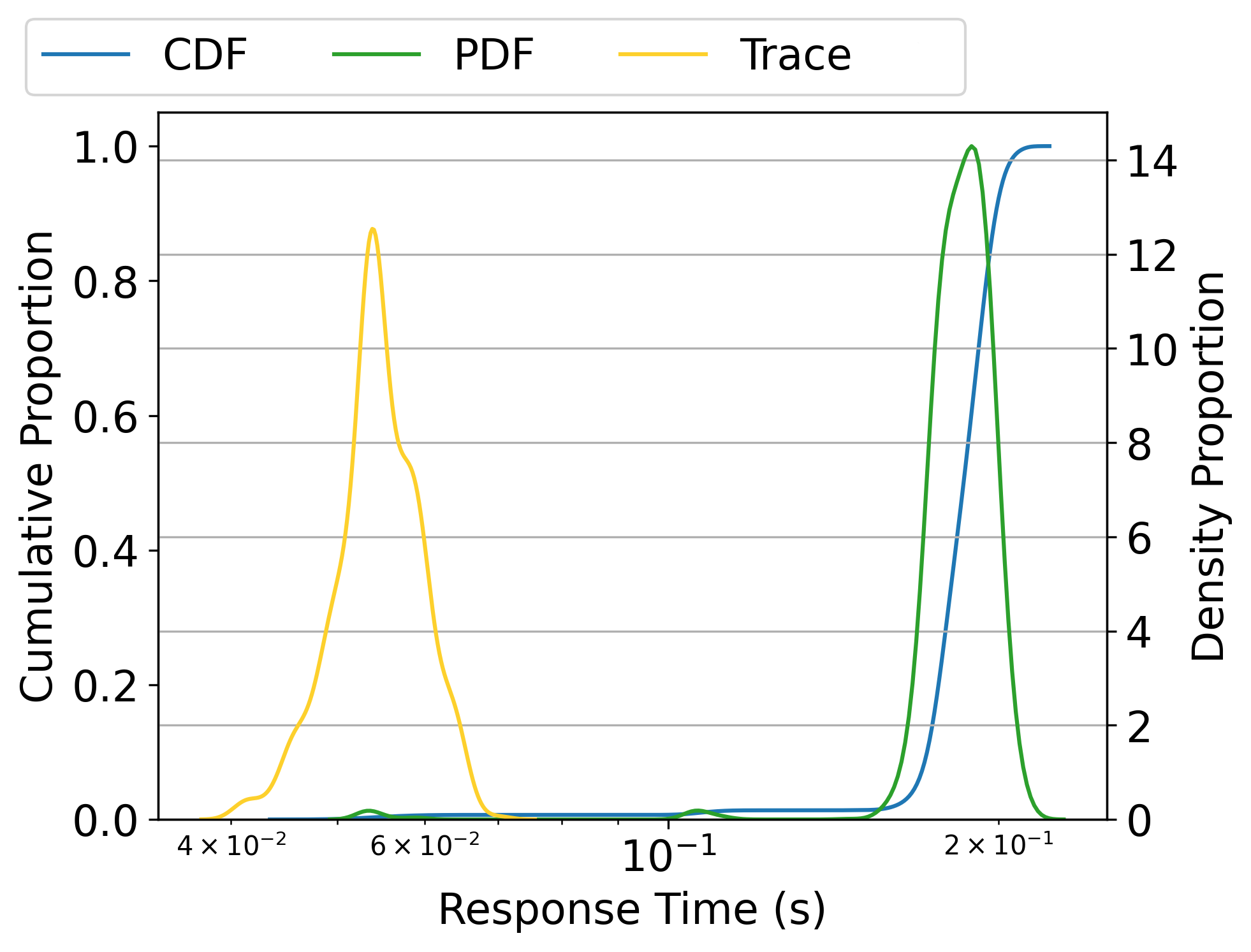}
    \caption{\gls{dp}}
    \label{fig:combined-cdf-subfig1}
  \end{subfigure}
  \hfill
  \begin{subfigure}[t]{0.32\linewidth}
    \centering
    \includegraphics[width=\linewidth]{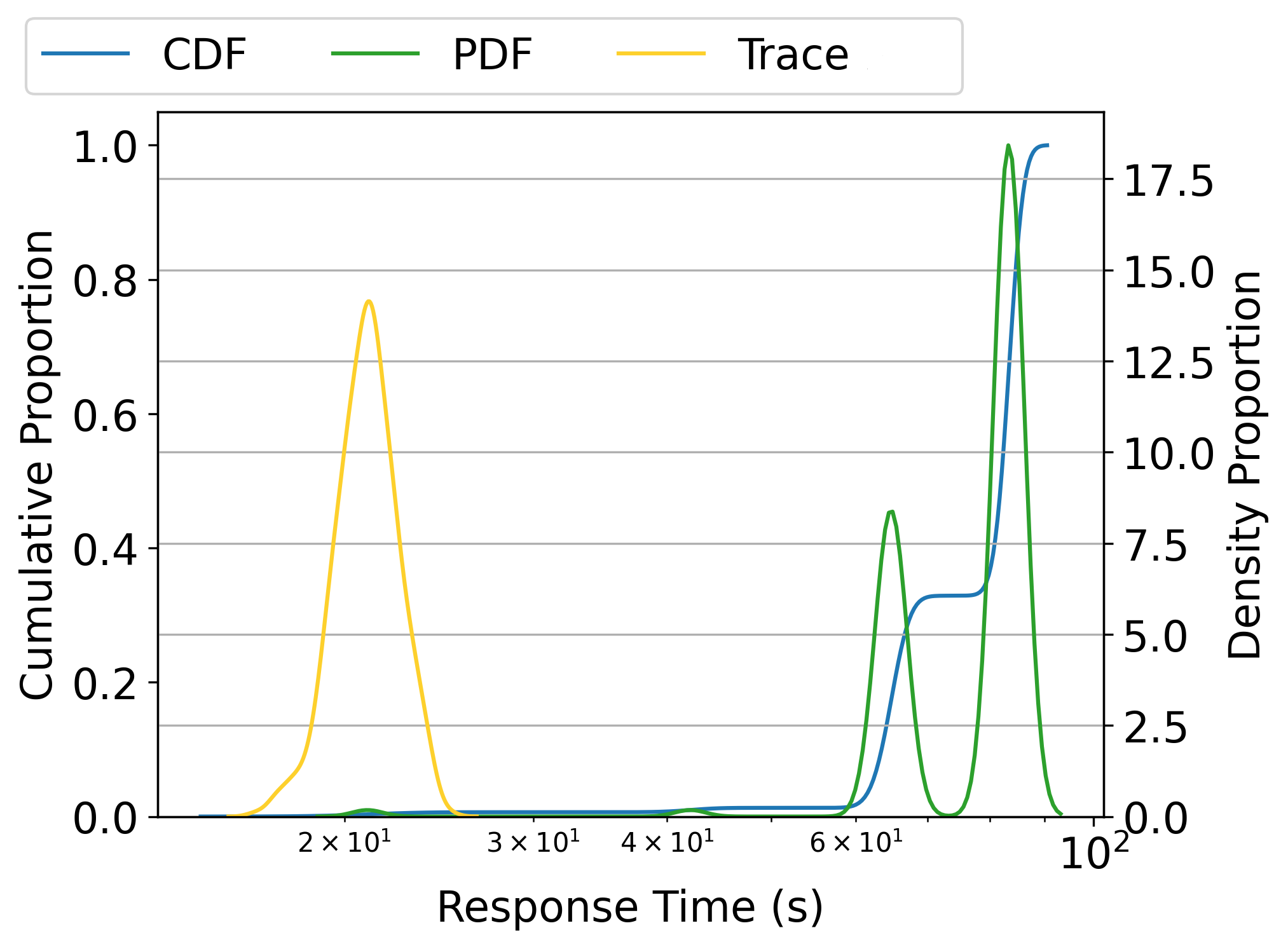}
    \caption{\gls{smc} 3-party in 500Mbps bandwidth}
    \label{fig:combined-cdf-subfig2}
  \end{subfigure}
  \hfill
  \begin{subfigure}[t]{0.32\linewidth}
    \centering
    \includegraphics[width=\linewidth]{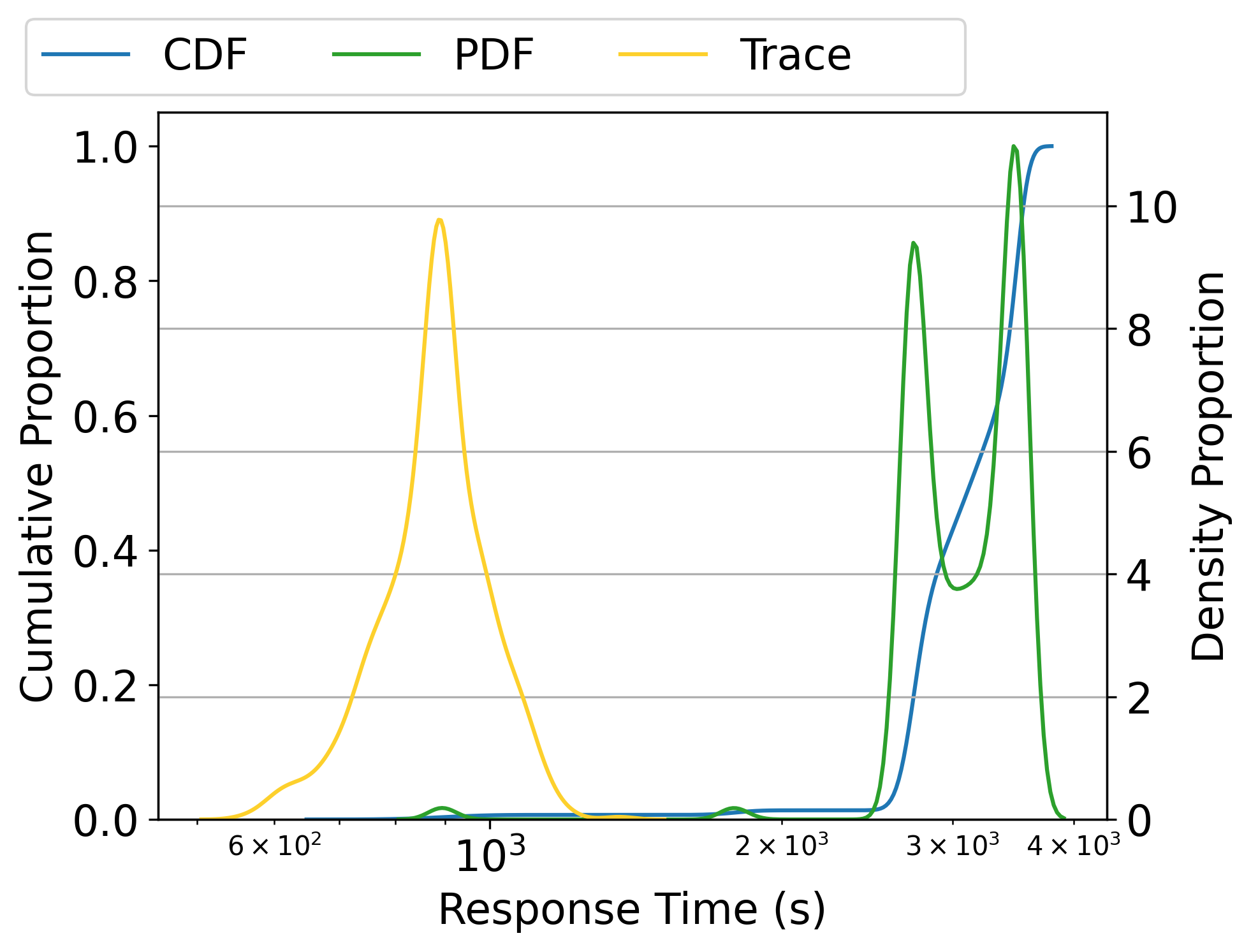}
    \caption{\gls{fhe}}
    \label{fig:combined-cdf-subfig3}
  \end{subfigure}
  \vspace{10pt}  

  \caption{PDF-CDF of response times for three techniques in AlexNet model on ECG5000 with 70 concurrent users}
  \label{fig:combined-cdf}
\end{figure*}

We measure the performance of the system using the average response time of inference tasks using the trace-based simulation.
Figure \ref{fig:dp-avgres} presents the simulation results of applying the \gls{dp} technique to deep learning models across three time-series datasets. Since the inference time with \gls{dp} remains comparable to that of the raw model, minimal overhead is expected, even with a large number of concurrent users.

The results reveal that LeNet-5 generally achieves the lowest response times, especially as the number of concurrent users increases, making it well-suited for high-load environments. In contrast, AlexNet consistently exhibits the highest average response time across all datasets, likely due to its more complex architecture. For example, at 70 concurrent users, AlexNet's response time is approximately 60-80\% higher than SqueezeNet on the ElectricDevices dataset. A similar pattern is observed for the FordA dataset, where LeNet-5 and SqueezeNet maintain significantly lower response times compared to AlexNet, with LeNet-5 slightly outperforming SqueezeNet as concurrency increases.

Overall, the average response time of models running \gls{dp} increases with the number of simultaneous users across all datasets, indicating a strain on the system under higher user concurrency. LeNet-5 consistently shows the most efficient scaling, with a steady increase in response time and a smaller slope relative to the other models. In contrast, AlexNet experiences the steepest increase, reflecting its higher computational load due to architectural complexity. This analysis highlights LeNet-5's suitability for real-time applications, particularly in scenarios with multiple concurrent users.

Regarding the simulation results for \gls{fhe}, Figure \ref{fig:concrete-avgres} shows the average response times for various models and datasets with increasing numbers of concurrent users. While the average response time grows by approximately 40\% for every 10 additional users, there is a significant gap in performance between LeNet-5, SqueezeNet, and AlexNet. In the ECG5000 dataset, SqueezeNet's response time is about 8-9 times higher than LeNet-5 at 30 users, and this difference remains consistent as user load increases. In contrast, AlexNet’s response time is nearly 11 times higher than LeNet-5’s at 30 users, reflecting the impact of AlexNet’s complexity and resulting in considerably higher latency across all user counts. A similar trend is observed in the FordA dataset, where AlexNet’s response time is roughly 1.2 times that of SqueezeNet and 10 times that of LeNet-5 at 30 users, highlighting the significant computational demand of AlexNet under \gls{fhe}.

Overall, AlexNet consistently exhibits the highest response times across all datasets, with the FordA dataset showing the most extreme values (up to 10,000 seconds at 70 users). This increase in response time is primarily due to AlexNet’s complexity, which imposes a substantial computational load under \gls{fhe}. Additionally, response times generally increase with higher user concurrency, ranging from 150\% to 400\% across different models and datasets. LeNet-5 scales more efficiently with user load, while AlexNet experiences the steepest increase.

These results highlight the significant computational overhead introduced by \gls{fhe}, which is about a 1000-fold increase in response time compared to \gls{dp} results in Figure \ref{fig:dp-avgres}. This impact is particularly pronounced for complex models like AlexNet and challenging datasets such as FordA. In contrast, simpler models like LeNet-5 exhibit relatively better performance in encrypted environments, making them more suitable for applications requiring real-time performance.

Figures \ref{fig:subfig-two-party-250}, \ref{fig:subfig-two-party-500}, \ref{fig:subfig-three-party-250}, and \ref{fig:subfig-three-party-500} illustrate the results of simulating the \gls{smc} technique across various models with different time-series datasets in two-party and three-party configurations. As outlined in the experimental setup, the goal is to assess the impact of communication between parties in \gls{smc}; thus, experiments are conducted in both 250 Mbps and 500 Mbps bandwidth environments. 

At the two-party configuration in 250 Mbps bandwidth, LeNet-5 on ECG5000 shows the lowest response times, from 0.7 seconds at 30 users and increasing to 1.2 seconds at 70 users. AlexNet on FordA is the slowest, reaching about 100 seconds at 70 users. This is over 8,000\% higher than LeNet-5 on ECG5000, highlighting the significant computational load of AlexNet. 
The response times in 500 Mbps configuration decrease across all models compared to that of 250 Mbps. For example, AlexNet on FordA in a two-party setting remains the slowest, reaching approximately 70 seconds at 70 users, a 30\% reduction from the 250 Mbps setting but still much higher than the other models.

Regarding the three-party configuration in Figures \ref{fig:subfig-three-party-250} and \ref{fig:subfig-three-party-500}, SqueezeNet running on ElectricDevices in 250 Mbps bandwidth sees an increase, with response times around 4 seconds at 30 users, increasing to about 8 seconds at 70 users, representing a 100\% increase compared to two-party settings. In 500 Mbps bandwidth, the model performance reduced to around 2 seconds at 30 users, rising to 4 seconds at 70 users, about a 33\% reduction from 250 Mbps.

Overall, the results demonstrate the significant overhead introduced by applying \gls{smc} compared to \gls{dp}, with a 9-fold increase in response time for the two-party configuration. While the two-party setup consistently achieves lower average response times than the three-party configuration across all datasets and bandwidths, doubling the bandwidth from 250 Mbps to 500 Mbps generally reduces response times for both configurations. This improvement is particularly evident for more complex models like AlexNet. Notably, the impact of increased bandwidth is more pronounced in the three-party setting, where response times at 250 Mbps are significantly higher than at 500 Mbps. These findings underscore the importance of carefully selecting the optimal setup to balance computational and communication overhead.


Additionally, Figure \ref{fig:combined-cdf} illustrates the cumulative distribution function (CDF) and the probability density function (PDF) of response times for the simulation experiment and implementation for the AlexNet model with 70 concurrent users on the ECG5000 dataset. It provides a detailed analysis of response time distributions across all three techniques. The PDF curves in each subplot (\ref{fig:combined-cdf-subfig1}, \ref{fig:combined-cdf-subfig2}, and \ref{fig:combined-cdf-subfig3}) resemble mixtures of Gaussian distributions, with multiple peaks indicating distinct response time clusters. Each subplot also includes the PDF of inference time traces obtained from the hardware implementation, offering additional insight into real-world performance variations. Among the techniques, \gls{dp} exhibits the least variability, forming a nearly Gaussian distribution similar to the trace distribution, with average response times of 0.0452s and 0.185s, respectively. In contrast, \gls{smc} shows moderate variability, likely influenced by network conditions, multi-party interaction, and resource allocation, resulting in a bimodal distribution. \gls{fhe} has the highest variability, with multiple peaks suggesting diverse response time clusters, primarily due to the computational overhead of homomorphic encryption.

Also, Table \ref{table:avg-response} shows the average response time breakdown for three privacy techniques on the AlexNet model with 70 concurrent users. The queuing time was measured during the simulation experiment, but the other components are based on the collected traces. \gls{dp} is the fastest (0.185s) due to minimal processing overhead, making it ideal for real-time use. \gls{fhe} is significantly time consuming (3102.03s), with massive inference time (663.91s) and queuing time (2433.16s). \gls{smc} offers a middle ground (76.775s), but still suffers from significant inference time (11.403s) and communication (9.6s) overhead.

\begin{table*}[ht]
\centering
\scriptsize
\caption{Average response time breakdown for AlexNet model on ECG5000 with 70 concurrent users.}
\renewcommand{\arraystretch}{1.5} 
\begin{tabular}{|l|c|l|c|}
\hline
\diagbox{\textbf{Time (s)}}{\textbf{Technique}} & \textbf{DP}   &\textbf{SMC - 3-party - 500Mbps}  & \textbf{FHE}    \\ \hline
Data Pre-processing                     & 0.001    &N/A   & N/A   \\ \hline
Data/Model Encryption  & N/A    &0.147   & 4.960   \\ \hline
Inference Time & 0.0442    &11.403   & 663.91   \\ \hline
Communication & N/A &9.6& N/A\\ \hline
Queuing Time &  0.1398 &55.624& 2433.16\\ \hline
Total & 0.185 &76.775& 3102.03\\ \hline
\end{tabular}
\label{table:avg-response}
\end{table*}

While \gls{dp} achieves performance comparable to the raw model, \gls{fhe} and \gls{smc} exhibit notable performance trade-offs. \gls{fhe}’s response time is heavily influenced by model complexity, leading to significant delays, whereas \gls{smc}’s performance is primarily constrained by communication overhead and the number of parties involved.

\subsection{Energy Consumption}

We measure the average energy consumption per inference for each model for ECG5000 dataset with different techniques. 
As shown in Table \ref{table:avg-energy}, Concrete-ML running \gls{fhe} is the most energy-intensive technique, especially for complex models. The energy consumption reaches 20.0167 Wh for AlexNet, making FHE a costly choice for high-accuracy, privacy-preserving machine learning on resource-constrained devices. The experiment with \gls{smc} ranks second to \gls{fhe} in terms of energy intensity. \gls{smc} energy consumption increases with model complexity and is affected by network bandwidth, with faster connections resulting in lower energy usage. For example, a 2-party SMC with AlexNet at 250 Mbps consumes 0.4766 Wh, while at 500 Mbps, it decreases to 0.3093 Wh. Increasing from 2-party to 3-party SMC further raises energy requirements, indicating that each additional participant introduces a small but measurable increase in resource usage. The third technique, \gls{dp}, consistently yields the lowest energy consumption across all models (0.0009–0.0011 Wh), demonstrating high energy efficiency. Energy usage also scales with model complexity, AlexNet consumes the most, while simpler models like LeNet-5 require less.

\begin{table*}[ht]
\centering
\scriptsize
\caption{Average energy consumption in watt-hours (Wh) per inference for different techniques running on ECG5000.}
\renewcommand{\arraystretch}{1.5} 
\begin{tabular}{|c|c|c|c|}
\hline
\diagbox{\textbf{Technique}}{\textbf{Model}} & \textbf{LeNet-5}  & \textbf{SqueezeNet}    & \textbf{AlexNet}  \\ \hline
DP                      & 0.0009   & 0.0009   & 0.0011   \\ \hline\hline
SMC - 2-party - 250Mbps  & 0.0111   & 0.0417   & 0.4766   \\ \hline
SMC - 2-party - 500Mbps  & 0.0088   & 0.0296   & 0.3093   \\ \hline
SMC - 3-party - 250Mbps  & 0.0186& 0.0713& 0.7436\\ \hline
SMC - 3-party - 500Mbps  & 0.0158& 0.0519& 0.4997\\ \hline\hline
FHE                      & 1.1168   & 13.1325  & 20.0167  \\ \hline
\end{tabular}
\label{table:avg-energy}
\end{table*}

\subsection{Discussion}
\label{section:discussion}

Selecting the right privacy-preserving technique is essential for ensuring both performance and privacy for \gls{ei} applications. This decision involves balancing two key considerations: the required level of privacy and the distributed nature of edge infrastructure. In particular, applications such as smart healthcare and autonomous vehicles demand not only strong data protection but also low-latency and high-reliability processing \cite{rancea2024edge}. In these edge intelligence scenarios, \gls{ppml} plays a critical role in achieving this balance, enabling secure, real-time inference while adapting to the constraints of edge devices and networks.

In smart healthcare, devices like wearable monitors or hospital-edge servers must process sensitive data in real time, for instance, to detect early signs of cardiac events or predict sepsis \cite{ochoa2024efficacy, qiu2020smartwatch}. Here, \gls{dp} can be employed during model training via DP-SGD to ensure patient-level privacy without exposing raw medical records. However, this approach introduces noise that may slightly reduce model accuracy, necessitating a careful trade-off between diagnostic precision and privacy guarantees. For collaborative inference involving multiple devices on sensitive physiological data, \gls{smc} combined with TLS ensures that patient data remains encrypted throughout the inference process and during transmission to edge servers. This setup maintains strong privacy protection even when operating over unsecured networks, enabling secure real-time diagnostics without exposing raw patient data.

In autonomous vehicles, low-latency decision-making is crucial, for example, real-time object detection, obstacle avoidance, or vehicle-to-vehicle coordination \cite{mao20233d}. \gls{fhe} enables vehicles to process encrypted sensor inputs (like LiDAR or camera data) on edge servers without ever decrypting them, thus protecting in-transit data. Due to \gls{fhe}’s computational demands in Concrete-ML, systems must be optimized by reducing model complexity, quantizing models (e.g., to 4–6 bits), and offloading computation to powerful edge servers. These optimizations ensure encrypted inference can meet strict latency requirements (e.g., less than 50 milliseconds) while keeping energy consumption within acceptable limits (e.g., around 5 Watt per vehicle). Similarly, in \gls{smc}-based collaborative perception tasks, reducing the number of participants and improving network bandwidth can significantly lower latency, allowing timely navigation decisions.

Our findings indicate that the system-level performance of \gls{ppml} techniques is shaped by both the underlying communication infrastructure and the scale of deployment. For example, \gls{smc} protocols degrade as party counts and network delays increase, whereas \gls{fhe} is bounded primarily by compute availability. Furthermore, while Concrete-ML automates the process of compiling models, it does not encrypt or secure the model itself. As a result, careful quantization and parameter tuning are essential to maintain both inference accuracy and energy efficiency. Across these applications, privacy methods must be selected and configured not in isolation, but in the context of real-world requirements such as detection accuracy, latency thresholds, energy budgets, and regulatory constraints. By grounding \gls{ppml} deployment in specific edge-intelligent use cases, designers can achieve secure, efficient, and responsive systems that uphold both functional and ethical standards.


\section{Security Analysis}
\label{section:security-analysis}

We analyze the protection goals, threat models, and practical leak channels for three families of privacy technologies used in our edge inference pipeline: \textbf{(i)} \gls{dp} with TensorFlow Privacy during training, \textbf{(ii)} \gls{smc} with CrypTen for encrypted inference, and \textbf{(iii)} \gls{fhe} with Concrete-ML for encrypted inference. Table~\ref{tab:privacy_analysis} summarizes the key aspects of this security analysis for each technique.

\begin{table*}[t]
\centering
\scriptsize
\renewcommand{\arraystretch}{1.5} 
\caption{Privacy focus, main knobs, and extraction impact.}
\label{tab:privacy_analysis}
\begin{tabular}{| p{0.17\linewidth} | p{0.26\linewidth} | p{0.27\linewidth} | p{0.24\linewidth} |}
\hline
\textbf{Method} & \textbf{Privacy Focus} & \textbf{Primary Knobs} & \textbf{Extraction Impact} \\ \hline
DP (TensorFlow Privacy) & Training-data privacy; $(\varepsilon,\delta)$-DP & Noise multiplier $\sigma$, clipping, sampling rate; accountant $\Rightarrow (\varepsilon,\delta)$ & Outputs unchanged; higher $\sigma$ typically increases $\Delta$-metrics (harder extraction) but reduces target accuracy \\ \hline
SMC (CrypTen) & Inference-time confidentiality among parties & \#parties, 16-bit fixed-point precision, ring $\mathbb{Z}_{2^{64}}$, Beaver triples & Predictions (in clear to querier) comparable to plaintext; extraction feasibility similar to non-private unless outputs restricted \\ \hline
FHE (Concrete-ML) & Client input confidentiality under 128-bit security & Security level (e.g., 128-bit), quantization, circuit depth & Predictions decrypted to client; extraction similar to non-private unless outputs perturbed/access-controlled \\ \hline
\end{tabular}
\end{table*}

\subsection{DP with TensorFlow Privacy}
\label{subsec:dp}

\paragraph{Threat model and guarantee.}
\gls{dp} bounds what can be inferred about any single training example from the learned model. A randomized mechanism $\mathcal{M}$ is $(\varepsilon,\delta)$\nobreakdash-DP if for all adjacent datasets $D,D'$ differing in one record and all measurable sets $S$ \cite{RN250, dwork2014algorithmic},
\begin{equation}
\Pr[\mathcal{M}(D)\in S] \le e^{\varepsilon}\Pr[\mathcal{M}(D')\in S] + \delta.
\end{equation}
Smaller $\varepsilon$ (for fixed $\delta$) implies stronger privacy. 
The parameter \(\delta\) is a failure probability: with probability at most \(\delta\), the mechanism may deviate from the pure \(\varepsilon\)-DP bound. To make such events vanishingly unlikely, \(\delta\) is chosen negligible in the dataset size. In our setting, we fix \(\delta\) to a negligible rate, e.g., \(\delta \le N^{-(1+\gamma)}\), so the overall guarantee holds except with probability at most \(\delta\). Here \(N\) denotes the size of the private training dataset (the number of distinct training records that incur privacy loss).
We train with DP\mbox{-}SGD: per\mbox{-}example gradients are $\ell_2$-clipped and Gaussian noise $\mathcal{N}(0,\sigma^2C^2I)$ is added to the clipped sum each step. A privacy accountant based on \gls{ma} or \gls{rdp} composes the per\mbox{-}step guarantees to an overall $(\varepsilon,\delta)$ given the noise multiplier $\sigma$, batch size $B$, sampling rate $q{=}B/N$, and number of steps $T$ (epochs $\approx T \cdot B/N$). 

\paragraph{Security knobs and side effects.}
Increasing $\sigma$ lowers $\varepsilon$ (stronger privacy) but typically reduces utility (accuracy). DP constrains \emph{training-time} leakage and provides robustness against inference that targets the inclusion or attributes of specific training points; however, it does not cryptographically hide inference inputs/parameters and does not, by itself, prevent black-box model extraction from released predictions.

\paragraph{Attacks mitigated vs.\ not mitigated.}
\gls{dp} directly mitigates training-data inference attacks such as membership inference, attribute inference about individual training records, and certain property/linkage tests that rely on overfitting to specific examples. By contrast, DP does not hide the functional behavior of the deployed model: a black-box API can still be queried to distill a surrogate (model stealing), unless outputs are restricted or perturbed at inference time \cite{liu2021generalization, zhao2025systematic}. In our evaluation (Section~\ref{section:dp-extraction-eval}), we quantify this tension by sweeping $\sigma$, reporting the induced $(\varepsilon,\delta)$, and measuring how the extraction gap $\Delta$ changes with the attacker’s query budget.

\subsection{SMC with CrypTen}
\label{subsec:smc}

\paragraph{Threat model and guarantee.}
SMC protects \emph{data in use} during inference against semi-honest (honest-but-curious) parties \cite{RN241}. In CrypTen, inputs and model parameters are secret-shared and the network is evaluated over an arithmetic ring without revealing raw values to any single party. The default instantiation provides information-theoretic privacy against up to \emph{one} corrupted party (i.e., $t{=}1<n/2$) under the semi-honest model. Malicious-secure variants require additional checks (e.g., MACs/consistency proofs) and are not enabled by default.

\paragraph{Implementation characteristics.}
The communication layer must use authenticated encryption, such as TLS, which is supported by CrypTen, to prevent traffic inspection, tampering, and session hijacking that could otherwise leak metadata or enable active disruption \cite{knott2021crypten}. Trust in the execution platform is likewise foundational: each party should run on an isolated, hardened node (VMs or container sandboxes with least privilege), because a compromised host can capture plaintext at ingress/egress or instrument the runtime to exfiltrate shares. Side channels remain a residual risk: variable batch sizes, data-dependent control flow, and non-constant-time kernels can correlate timing, message sizes, or round counts with private inputs. Finally, code provenance and binary integrity are part of the threat surface, if an adversary can alter delivered artifacts or dependencies, they can introduce covert disclosure while preserving protocol transcripts.

\paragraph{Security knobs and side effects.}
Our results show that increasing the number of parties primarily impacts communication, latency, and round complexity, but does not change model accuracy, since the secure protocol computes the same function. In contrast, precision/quantization settings (e.g., fixed-point fractional bits under the default 16-bit encoding and the $\mathbb{Z}_{2^{64}}$ ring) directly affect numerical error and thus can impact accuracy: too few fractional bits or poor scaling increase quantization error; overly aggressive scaling risks modular wraparound. SMC does not degrade performance like DP; accuracy typically matches plaintext given adequate precision. SMC does not prevent model extraction when an adversary can query the black-box and observe cleartext outputs, since the confidentiality guarantee covers inputs/parameters during computation, not the information content of released predictions.
Reducing precision (fewer fractional bits) can substantially improve throughput and bandwidth, but the trade is purely \emph{numerical}: it may lower task accuracy or increase wraparound risk; it does not reduce the $t{=}1$ semi-honest corruption threshold or the information-theoretic privacy of shares. Conversely, moving from semi-honest to malicious security strengthens guarantees (detecting active tampering) at additional compute/communication cost.

\subsection{FHE with Concrete-ML}
\label{subsec:fhe}

\paragraph{Threat model and guarantee.}
FHE protects \emph{data in use} by enabling computation directly on encrypted inputs (and optionally encrypted model parameters depending on the deployment pattern). Concrete-ML uses leveled/FHE circuits (with programmable bootstrapping) and by default targets at least 128-bit classical security \cite{ZamaConcreteML}.

\paragraph{Implementation characteristics.}
Models are quantized (e.g., to 4-bit integers) and compiled to FHE circuits; inference returns encrypted outputs that the client decrypts. As with SMC, FHE keeps inputs hidden from the model owner and infrastructure.

\paragraph{Security knobs and side effects.}
Security level (e.g., 128-bit) is set by lattice parameters; increasing precision or circuit depth increases latency. Like SMC, FHE \emph{does not} inherently mitigate model extraction from accessible (decrypted) outputs. With sufficient numeric precision (e.g., appropriate quantization scale and polynomial approximations), its predictive accuracy typically matches the plaintext model. Any residual loss is due to implementation limits (precision, polynomial depth, compiler constraints).



\subsection{Privacy under Model Stealing}
\label{section:dp-extraction-eval}

A model–stealing (model–extraction) attack treats the deployed predictor as a black box: the adversary collects or synthesizes inputs \(x\), queries the API, records the returned predictions, and trains a surrogate to mimic the target’s decision function. This is prominent in \gls{ei} deployments, where prediction endpoints are exposed to field devices: even without access to weights or training data, a compromised device can generate realistic query traffic and harvest outputs. We adapt a protocol previously examined for \gls{fhe} in cloud settings~\cite{balaban2025privacy}, but tailor it to \gls{ei}: rather than presuming the attacker holds a subset of the training data, we assume the adversary compromises an edge device, harvests sensor traces, and issues black–box queries to the remotely hosted target model at the edge server layer, observing only top–1 outputs. This captures the operational reality of \gls{ei}, where training data are centralized and inaccessible, but field devices can be subverted.

We concentrate our empirical study on \gls{dp} rather than \gls{smc} or \gls{fhe} because DP changes the learned model in a way that can affect extractability. DP-SGD clips per-example gradients and adds Gaussian noise. This lowers target accuracy and smooths decision boundaries; with a fixed query budget, the surrogate lags further behind, revealing a privacy--utility--stealability trade-off controlled by \(\sigma\). In contrast, \gls{smc} (CrypTen) and \gls{fhe} (Concrete-ML) cryptographically protect inputs, parameters, and the evaluation procedure, but once predictions are revealed their information content matches plaintext unless extra defenses (output perturbation, throttling, access control) are applied, so stealing curves would resemble plaintext, only slower to run. Practically, DP (via TensorFlow Privacy) is easy to sweep over \(\sigma\), letting us report both attack outcomes and the corresponding \(\varepsilon\). Accordingly, we quantify how DP training, via a \(\sigma\) sweep, affects the attacker’s ability to approach the target under fixed query budgets in our \gls{ei} threat model.

\subsubsection{Attack Model and Protocol}
\label{subsec:attack-model}

We consider a black-box adversary that can submit inputs to the target and observe top-1 predictions (no white-box access). The attacker selects a fraction $q$ of the held-out test set as queries and augments them with synthetic samples generated by Gaussian perturbations around queried points. The DP-trained target is queried on this union to obtain stolen labels $\hat{y}(t)$. A non-private substitute model is then trained with cross-entropy on the attacker dataset (real+synthetic, stolen labels) using a validation split for early selection. Finally, both the DP target and the substitute model are evaluated on the disjoint $(1-q)$ holdout.

We report macro-averaged Accuracy, Precision, Sensitivity (Recall), and F$_1$. In addition, following security-centric practice, we measure the performance gaps
\begin{equation}
\begin{aligned}
\Delta\text{Acc}  &= \text{Acc}_{\text{target}} - \text{Acc}_{\text{sub}}, \\
\Delta\text{Prec} &= \text{Prec}_{\text{target}} - \text{Prec}_{\text{sub}}, \\
\Delta\text{Sens} &= \text{Sens}_{\text{target}} - \text{Sens}_{\text{sub}}, \\
\Delta\text{F$_1$}   &= \text{F$_1$}_{\text{target}}   - \text{F$_1$}_{\text{sub}}.
\end{aligned}
\end{equation}
where larger gaps indicate the attacker is further from the target (harder extraction).

\subsubsection{Implementation}
\label{subsec:impl}
We evaluate on the FordA time\mbox{-}series dataset. The \emph{target} is trained with TensorFlow Privacy applying DP\mbox{-}SGD (per\mbox{-}example clipping and Gaussian noise) while sweeping the noise multiplier $\sigma\!\in\!\{0.1,0.2,\ldots,1.0,2.0\}$; A standard privacy accountant reports the corresponding $(\varepsilon,\delta)$ at the end of training. For the \emph{attacker}, we sample a fraction $q\!\in\!\{0.15,0.30\}$ of the test set as the query pool, then generate synthetic queries by resampling with replacement and adding IID Gaussian jitter (noise scale $0.05$). We set the synthetic factor to $1.0$, producing as many synthetic points as real ones, and query the black\mbox{-}box target on the union of real and synthetic inputs to obtain stolen labels.

The \emph{substitute} is a plain LeNet trained with cross\mbox{-}entropy on the stolen labels; we select the best checkpoint by validation accuracy on a held\mbox{-}out split of the attacker’s data. Evaluation uses the disjoint $(1\!-\!q)$ holdout to compute macro Accuracy, Precision, Sensitivity (Recall), and F$_1$ for both target and substitute, and we log per\mbox{-}run JSON (including $\Delta$-metrics, Target$-$Substitute). All experiments are repeated with $5$ random seeds; we report mean$\pm$std for every metric across seeds and across the $\sigma$ sweep.

For each $\sigma$, we also compute the privacy loss $\varepsilon (\sigma;\allowbreak N,\allowbreak B,\allowbreak T,\allowbreak \delta)$ using an \gls{rdp}-based accountant. We then plot the $\Delta$-metrics against $\sigma$ \emph{and} overlay (or co-plot) the corresponding $\varepsilon$ values, enabling a joint view of \emph{utility/extractability} (via $\Delta$-metrics) versus \emph{formal privacy} (via $\varepsilon$). This makes it easy to identify operating points where $\varepsilon$ is small (stronger privacy) while $\Delta\text{Acc}$ remains large (the substitute lags the target).

\subsubsection{Results Summary}
\label{subsec:results}

\begin{figure}[t]
  \centering
  \includegraphics[width=\linewidth]{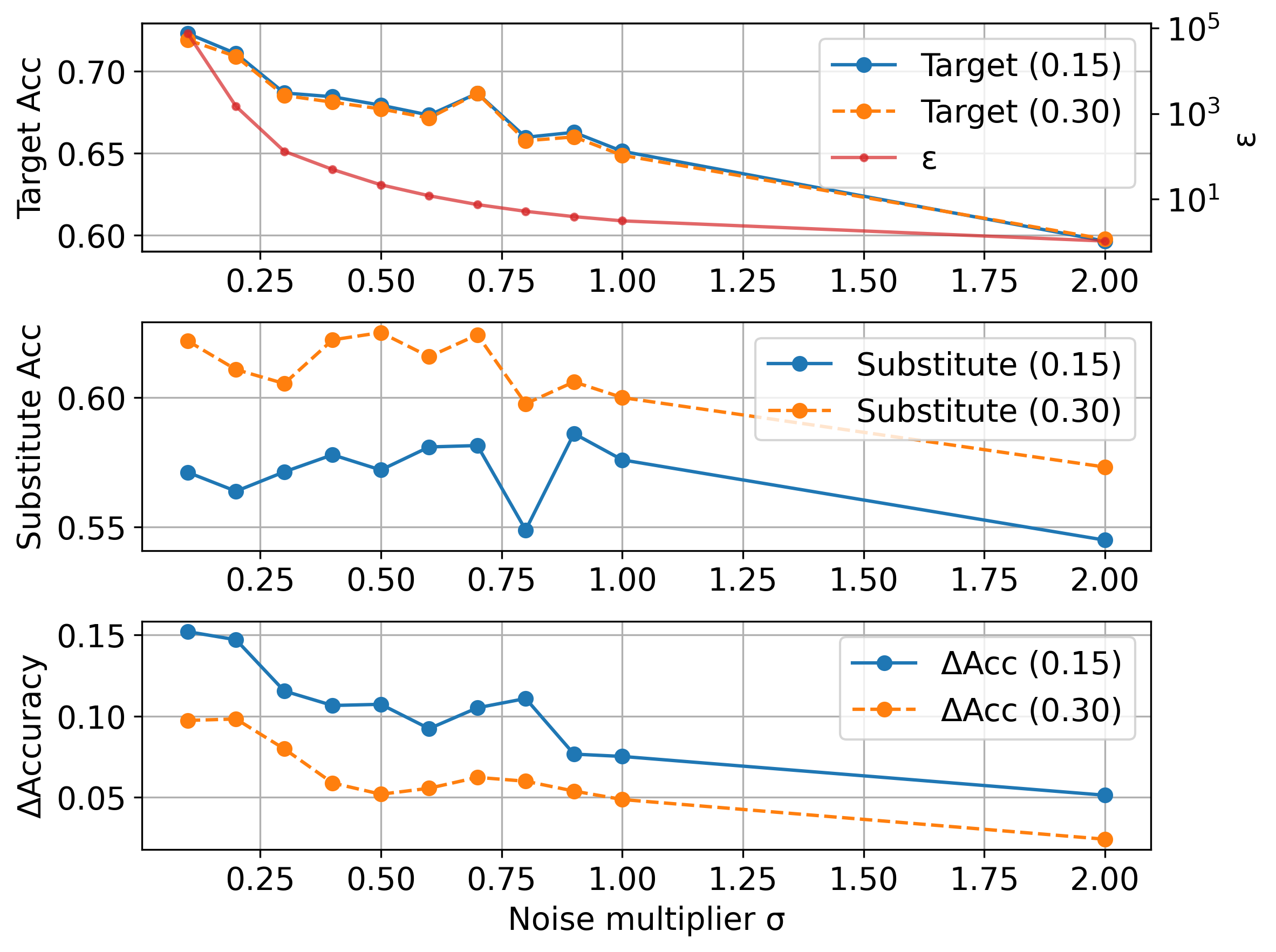}
  \vspace{0.35em}
  \caption{Differential Privacy vs.\ model stealing on \textit{FordA} with LeNet.
  We vary the DP noise multiplier $\sigma$ and compare two attacker query-pool fractions
  $q\!\in\!\{0.15,0.30\}$. \emph{Top:} target accuracy; \emph{middle:} substitute
  accuracy; \emph{bottom:} accuracy gap $\Delta\mathrm{Acc}$.}
  \label{fig:dp_stealing_forda_lenet}
\end{figure}

Figure~\ref{fig:dp_stealing_forda_lenet} shows three coupled trends. First, as $\sigma$ increases (i.e., privacy strengthens), the target’s accuracy falls roughly monotonically: for $q{=}0.15$ it drops from about $72\%$ at $\sigma{=}0.1$ to about $60\%$ at $\sigma{=}2.0$. Second, the substitute’s accuracy is governed mainly by the attack budget: the $q{=}0.30$ curve sits consistently above $q{=}0.15$ (e.g., around $62\%$ vs.\ $57\%$ at small $\sigma$), reflecting that more queries plus synthetic augmentation yield a better surrogate. Consequently, the gap $\Delta\mathrm{Acc}$ narrows both when $q$ increases (stronger extractability) and when $\sigma$ becomes very large (the target itself weakens). In the mid-noise regime, the target remains viable while the attacker’s gains are modest, producing a larger and more desirable $\Delta\mathrm{Acc}$ (about $0.10$ near $\sigma{\approx}0.5$ for $q{=}0.15$, tapering to $\sim\!0.05$ by $\sigma{=}2.0$).

The overlaid $\varepsilon$ curve (top panel, right log axis) quantifies privacy as a function of $\sigma$ for FordA with training size $N{=}3601$, batch size $50$, $25$ epochs, and $\delta{=}1/N$. The accountant yields a precipitous drop in $\varepsilon$ as $\sigma$ grows: at very small noise ($\sigma{=}0.1$), $\varepsilon$ is extremely large (order $10^4$–$10^5$), offering little practical privacy; by $\sigma{\approx}0.5$, $\varepsilon$ is already well below $10$, and around $\sigma{\approx}1.0$ it reaches the $O(1)$ regime (substantially stronger privacy). Thus, increasing $\sigma$ simultaneously \emph{reduces} extractability (keeping the substitute behind the target) and \emph{lowers} $\varepsilon$, but at the cost of target utility. Across seeds we observe low variance for the target and higher variance for the substitute, especially at $q{=}0.15$, indicating that DP training also makes the attack less stable when the adversary’s query budget is constrained.

Overall, the figure exposes a three-way trade-off: raising $\sigma$ strengthens formal privacy (lower $\varepsilon$) and dampens stealing success (larger $\Delta\mathrm{Acc}$ at moderate noise), while larger $q$ improves the attacker (shrinking $\Delta\mathrm{Acc}$) without affecting the target directly. The most defensible operating points lie where $\varepsilon$ is in the single-digits and the target still retains acceptable accuracy (mid-$\sigma$), keeping the attacker data-limited.

To make this concrete, Table~\ref{tab:dp_gaps_forda_lenet_q30} summarizes how the extraction gap (Target $-$ Substitute) varies with the DP noise multiplier $\sigma$ on \textit{FordA}/LeNet when the attacker controls 30\% of the test set for queries. The gaps are largest at low noise and shrink as $\sigma$ increases: $\Delta\text{Acc}$ is about $0.097$ at $\sigma{=}0.1$, around $0.052$ at $\sigma{=}0.5$, and falls to $0.024$ at $\sigma{=}2.0$. Precision, Sensitivity, and F$_1$ track accuracy closely (typically within $\approx 0.002$), indicating that DP noise affects the surrogate’s class-wise behavior uniformly rather than skewing a particular metric.

\begin{table}[ht]
\centering
\scriptsize
\caption{Gap metrics (Target $-$ Substitute) vs.\ DP noise multiplier $\sigma$ on \textit{FordA} with LeNet for attacker fraction $q=0.30$. Means over 5 seeds.}
\renewcommand{\arraystretch}{1.5} 
\begin{tabular}{|c|c|c|c|c|}
\hline
\textbf{$\sigma$} & \textbf{$\Delta$Acc} & \textbf{$\Delta$Prec} & \textbf{$\Delta$Sens} & \textbf{$\Delta$F$_1$} \\ \hline
0.1 & 0.097 & 0.098 & 0.098 & 0.099 \\ \hline
0.2 & 0.098 & 0.098 & 0.099 & 0.100 \\ \hline
0.3 & 0.080 & 0.080 & 0.081 & 0.085 \\ \hline
0.4 & 0.059 & 0.060 & 0.059 & 0.060 \\ \hline
0.5 & 0.052 & 0.053 & 0.051 & 0.052 \\ \hline
0.6 & 0.056 & 0.056 & 0.056 & 0.058 \\ \hline
0.7 & 0.062 & 0.063 & 0.062 & 0.063 \\ \hline
0.8 & 0.060 & 0.062 & 0.060 & 0.061 \\ \hline
0.9 & 0.054 & 0.053 & 0.054 & 0.054 \\ \hline
1.0 & 0.049 & 0.050 & 0.048 & 0.048 \\ \hline
2.0 & 0.024 & 0.026 & 0.024 & 0.024 \\ \hline
\end{tabular}
\label{tab:dp_gaps_forda_lenet_q30}
\end{table}




\section{Conclusion}
\label{section:conclusion}

Privacy is a key challenge in \gls{ei} applications, driven by concerns over sensitive data and its potential impacts. This paper focuses on evaluating different privacy-preserving techniques for \gls{ei} applications using trace-based simulation. We provide an overview of the primary privacy-preserving approaches and their applications in edge environments, followed by a detailed description of the framework’s architecture, privacy-preserving frameworks, the associated security analysis and the experimental setup. We also present a case study of a black-box model–stealing attack against \gls{dp}, including an implementation and quantitative results.
Our findings reveal that while \gls{dp} achieves performance close to the raw model, it suffers significant accuracy loss due to added noise, particularly in complex models like AlexNet. \gls{fhe} faces substantial computational overhead, requiring careful system design, such as reducing model complexity and offloading inference to high-performance edge servers. \gls{smc} performance is highly sensitive to network bandwidth and the number of participating parties, with increasing parties leading to greater communication overhead and performance degradation. In future work, we plan to work on optimizing these techniques and developing new task scheduling algorithms to improve their performance in edge computing environments.

\section*{Acknowledgment}

This work was supported in part by AerVision Technologies. 

\section*{Declarations}

\begin{itemize}
  \item \textbf{Funding:} This research was supported in part by AerVision Technologies. 

  \item \textbf{Conflict of interest/Competing interests:} The authors declare that they have no competing interests.

  \item \textbf{Ethics approval and consent to participate:} This study did not involve human participants or animals, and therefore ethics approval and consent to participate were not required.

  \item \textbf{Consent for publication:} Not applicable.

  \item \textbf{Data availability:} The datasets used in this study are publicly available.

  \item \textbf{Materials availability:} Not applicable. 

  \item \textbf{Code availability:} The code used to reproduce the experiments is available from the corresponding author upon reasonable request.

  \item \textbf{Author contribution:} Quoc Lap Trieu (QLT) conceived the study, designed the methodology, implemented the framework, conducted the experiments, analyzed the results, and drafted the manuscript.
  Bahman Javadi Jahantigh (BJJ) supervised the research, contributed to study design and interpretation of results, and revised the manuscript critically.
  Jim Basilakis (JB) contributed to the experimental design and evaluation methodology, assisted with interpretation of results, and revised the manuscript.
  All authors reviewed and approved the final manuscript.
\end{itemize}

\bibliography{guides}

\end{document}